\pdfoutput=1
\RequirePackage{ifpdf}
\documentclass{JINST}

\usepackage{caption}
\usepackage{subcaption}
\usepackage{xspace}
\usepackage{lineno}

\title{Triggers for displaced decays of long-lived neutral particles in the ATLAS detector}

\author{The ATLAS Collaboration}

\newcommand{\ECal}{ECal\xspace}
\newcommand{\HCal}{HCal\xspace}
\newcommand{\CaloRatio}{E_{\rm{HAD}}/E_{\rm{EM}}\xspace}
\newcommand{\LogCaloRatio}{\log_{10}(E_{\rm{HAD}}/E_{\rm{EM}})\xspace}
\newcommand{\rois}{RoIs\xspace}
\newcommand{\roi}{RoI\xspace}
\newcommand{\piv}{$\pi_{\rm{v}}$\xspace}

\newcommand{\online}{online\xspace}

\newcommand{\Endcap}{end-cap\xspace}
\newcommand{\Endcaps}{end-caps\xspace}
\newcommand{\pileup}{pile-up\xspace}
\newcommand{\pT}{p_{\rm{T}}\xspace}
\newcommand{\ET}{E_{\rm{T}}\xspace}

\abstract{
A set of three dedicated triggers designed to detect long-lived neutral particles decaying throughout the ATLAS detector to a pair of hadronic jets is described. The efficiencies of the triggers for selecting displaced decays as a function of the decay position are presented for simulated events. The effect of pile-up interactions on the trigger efficiencies and the dependence of the trigger rate on instantaneous luminosity during the 2012 data-taking period at the LHC are discussed.}

\begin{document}

\section{ Introduction}
Many extensions to the Standard Model (SM) include the production of neutral, weakly-coupled and long-lived particles that can decay to final states containing several hadronic jets. These long-lived particles occur in many models such as Gauge Mediated Supersymmetry Breaking (GSMB) \cite{SUSY}, the Minimal Supersymmetric Standard Model (MSSM) with R-parity violation \cite{MSSM}, inelastic dark matter \cite{DarkMatter} and Hidden Valley (HV) scenarios \cite{HV1,HV2,HV3}. Such neutral particles with a potentially long decay path present a trigger and reconstruction challenge for the LHC detectors in that they have no detector activity connecting them to the primary interaction point (IP). Depending on the kinematics of the production mechanism and the masses of the parent and long-lived objects, these events may be displaced in time from the events selected by the standard triggers. In this paper we discuss the signature-driven triggers developed by the ATLAS Collaboration to select these events. To evaluate these triggers we use as a benchmark an HV scenario, in which the SM is weakly coupled, by a heavy communicator particle, to a hidden sector that includes a neutral pseudoscalar pion, \piv. The \piv can have a long lifetime resulting in decays far from the IP. One of these triggers was successfully used for the recently published search for a light Higgs boson decaying to long-lived neutral particles \cite{HVAtlas}. 

Section 2 describes the ATLAS detector and trigger system, section 3 discusses the benchmark model and the decay signatures, sections 4--6 describe the long-lived neutral particle triggers, section 7 presents trigger performance on simulated events and section 8 discusses trigger performance in the 2012 data-taking period.

\section{The ATLAS detector}
ATLAS is a multi-purpose detector \cite{ATLASTDR} consisting of an inner tracking system (ID) embedded in a superconducting solenoid, electromagnetic (\ECal) and hadronic (\HCal) calorimeters and a muon spectrometer (MS) with three air-core toroidal magnetic fields. 
The ID consists of a silicon pixel detector, a silicon strip detector (semiconductor tracker, SCT) and a straw tube tracker (transition radiation tracker, TRT) and provides precision tracking of charged particles for \mbox{$\mid\eta\mid$~$\leq2.5$}.\footnote{The ATLAS reference system is a cartesian right-handed coordinate system, with the nominal collision point at the origin. The anti-clockwise beam direction defines the positive $z$-axis, with the $x$-axis pointing to the centre of the LHC ring. The angle $\phi$ defines the direction in the plane $(x, y)$ transverse to the beam. The pseudorapidity is given by $\eta=-\ln \tan(\theta/2)$, where the polar angle $\theta$ is taken with respect to the positive $z$ direction.} The ID extends from a radius, $r$, of 0.05\,m to 1.1\,m and in $|z|$ to 3.5\,m. 
The \ECal and \HCal systems cover the region $|\eta|\leq4.9$ and have a thickness of 9.7 interaction lengths at $\eta=0$. In the region $|\eta|\leq3.2$ electromagnetic calorimetry is provided by barrel and \Endcap high-granularity lead/liquid-argon (lead/LAr) electromagnetic calorimeters. An additional thin LAr presampler covering $|\eta|\leq1.8$ is used to correct for energy loss in the material upstream of the calorimeters. Hadronic calorimetry is provided by a steel/scintillator tile calorimeter that is segmented into three barrel structures for $|\eta|\leq1.7$ and two copper/LAr hadronic \Endcap calorimeters that extend the coverage to $|\eta|\leq3.2$. The solid angle coverage is completed up to $|\eta|\leq4.9$ with forward calorimeter modules optimized for both electromagnetic and hadronic measurements. The \ECal extends from $r$ of 1.5\,m to 2.0\,m in the barrel and from $|z|$ of 3.6\,m to 4.25\,m in the \Endcaps. The \HCal covers the region $r$ from 2.25\,m to 4.25\,m in the barrel and $|z|$ from 4.3\,m to 6.05\,m in the \Endcaps. The MS provides trigger and momentum measurements for charged particles entering the spectrometer. It consists of one barrel and two \Endcaps, each with 16 sectors, equipped with fast detectors for triggering and precision tracking chambers, monitored drift tubes (MDT) and cathode strip chambers (CSC). In the MS barrel region, tracks are measured in chambers arranged in three cylindrical layers around the beam axis; in the transition and \Endcap regions, the chambers are installed in planes perpendicular to the beam, also in three layers. An air-core toroidal field allows accurate charged particle reconstruction independent of the ID information. For triggering, three stations of resistive plate chambers (RPC) and thin gap chambers (TGC) are used in the MS barrel and \Endcaps, respectively. The first two RPC stations are located at a radius of 7\,m (large sectors\footnote{The barrel chamber system is subdivided into large (between the magnet coils) and small (inside the magnet coils) sectors.}) and 8\,m (small sectors) with a separation of 0.5\,m; the third station is at a radius of 9\,m (large sectors) and 10\,m (small sectors). In the \Endcaps the first TGC station is located at $|z|=13$\,m and the other two at 14\,m with a separation of 0.5\,m.

The trigger system has three levels~\cite{TRIG}. 
The first level (L1) is a hardware-based system using information from the calorimeter and the muon detectors. The second (L2) and third (Event Filter, EF) levels are software-based systems using information from all of the ATLAS detectors. Together, L2 and EF are called the High Level Trigger (HLT). 
The L1 trigger uses information based on relatively coarse data from the calorimeters and the muon trigger stations. The L1 thresholds are applied to transverse energy ($\ET$) for calorimeter-based triggers and transverse momentum ($\pT$) for muon triggers. The L1 trigger identifies Regions of Interest (\rois), which are ($\eta$,$\phi$) regions of the detector associated to a specific physics signature. 
\rois are widely used in the subsequent trigger levels to reduce the amount of data read from the detector readout buffers while accessing the full information for the most important part of the event.

The L1 muon trigger logic is implemented in similar ways for both the RPC and TGC systems. Each of the three planes of the RPC system and the two outermost planes of the TGC system consist of a doublet of independent detector layers while the first TGC plane contains three detector layers. A low-$\pT$ muon \roi is generated by requiring a coincidence of hits in at least three of the four layers of the two inner RPC planes for the barrel and of the two outer TGC planes for the \Endcaps. A high-$\pT$ muon RoI requires additional hits in at least one of the two layers of the outer RPC plane for the barrel and in two of the three layers of the innermost TGC layer for the \Endcaps. The muon \rois have a spatial extent of 0.2 in $\Delta\eta$ and $\Delta\phi$ in the MS barrel and 0.1 in the MS \Endcaps. Only the two highest-$\pT$ \rois per MS sector are used by the HLT.

The L1 calorimeter trigger provides the capability to search for trigger objects such as electrons, photons, $\tau$ leptons, jets and transverse energy based on information from the calorimeter elements within projective regions, called trigger towers. 
The trigger towers have a size of approximately 0.1 in $\Delta\eta$ and $\Delta\phi$ in the central part of the calorimeter, \mbox{$\mid\eta\mid$~$<2.5$}, and are larger and less uniform in the more forward region.

\section{Benchmark model and decay signature}
An HV model with a light Higgs mediator is used to evaluate the ATLAS detector response to highly displaced decays to hadronic jets. Three different Monte Carlo (MC) simulation samples are used for this study. In each sample, a Higgs boson is produced via gluon fusion and decays to a pair of long-lived \piv particles. Because the \piv is a pseudoscalar, it decays predominantly to heavy fermions, $b\overline{b}$, $c\overline{c}$ and $\tau^{+}\tau^{-}$ in the ratio 85:5:8, as a result of the helicity suppression of the low-mass fermion anti-fermion pairs.\footnote{Hidden sector scalars that couple to the SM via their mixings with the Higgs boson have Yukawa interactions with fermions and anti-fermions and these couplings are proportional to the fermion masses; therefore, these scalars decay in a similar way to the pseudoscalar \piv particles.}
\begin{table}[h!]
\begin{center}
\begin{tabular}{ccc}
\hline
Higgs boson mass [GeV] & \piv mass [GeV] & \piv mean proper lifetime [m] \\
\hline
100 & 25 & 0.125 \\
126 & 10 & 0.35 \\
140 & 20 & 0.63 \\
\hline
\end{tabular}
\caption{Mass and mean proper lifetime parameters for the simulated benchmark models.}
\label{tab:MCParameters}
\end{center}
\end{table}

The parameters used to generate the samples are summarized in table \ref{tab:MCParameters}. The mean proper lifetime values are chosen to maximize the number of simulated \piv decays for all of the ATLAS detectors. The masses are chosen to give a range of $\beta$ (speed relative to the speed of light in vacuum) distributions of the \piv particles in order to study trigger timing and correct bunch crossing identification. The events are generated at a center-of-mass energy of 8\,TeV using Pythia 8.165~\cite{PYTHIA8}. The MSTW 2008 leading order parameterization \cite{PDF} is used for the parton distribution functions in the proton. The mean value of $\langle\mu\rangle$ used in the simulation is 22 and the distribution follows the predicted conditions for the 2012 $\sqrt{s}=8$\,TeV data, which ranged in instantaneous luminosity from approximately 2 to $7\times10^{33}$\,cm$^{-2}$\,s$^{-1}$. The quantity $\mu$ is a measure of the average number of inelastic interactions per bunch crossing and $\langle\mu\rangle$, the average value over all proton bunches, gives the average number of expected proton-proton collisions per event.
\begin{figure}[t!]
	\centering
	\includegraphics[width=0.7\textwidth]{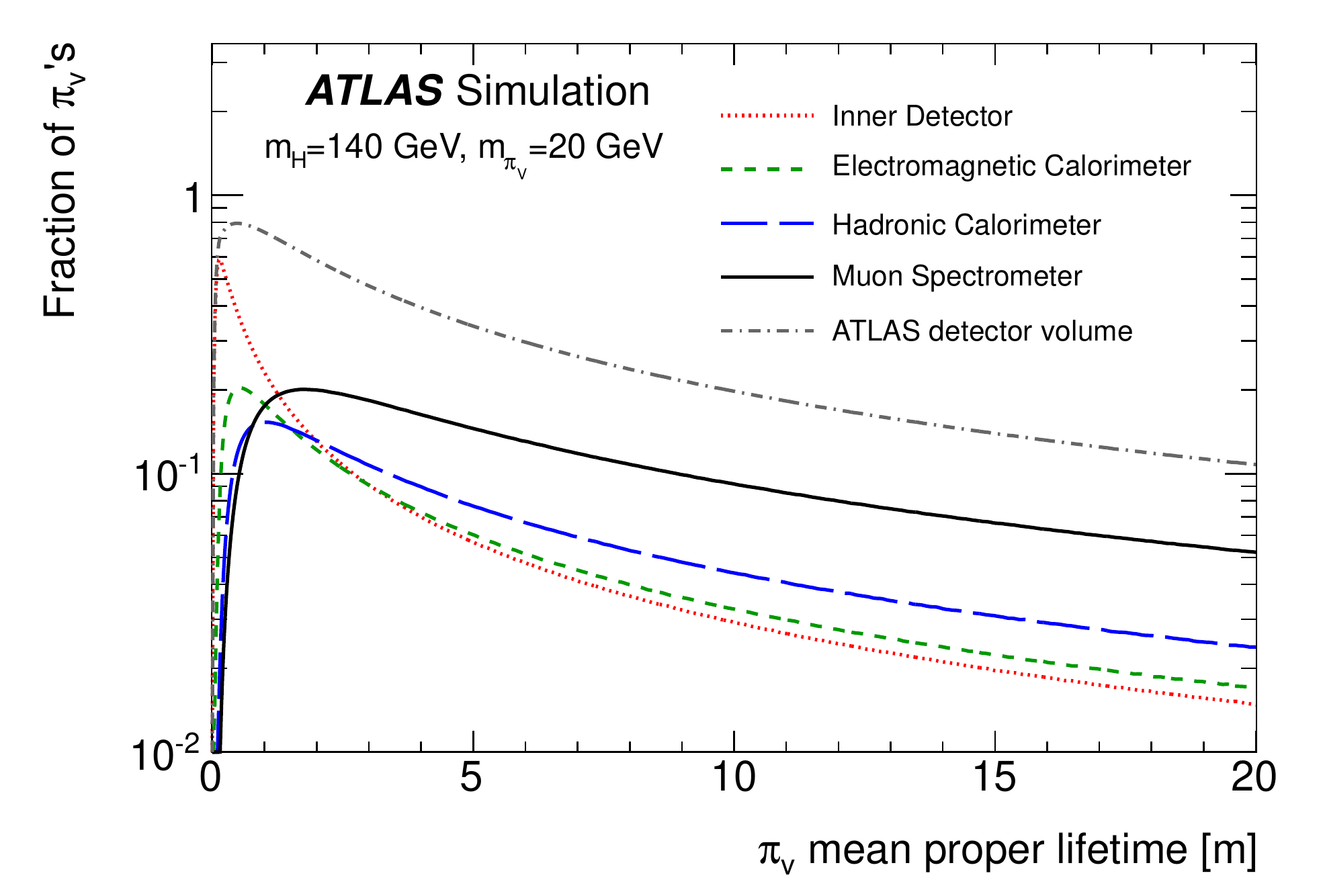}
	\caption{The probability of a \piv to decay in the ID (beyond the second layer of the pixel detector), \ECal, \HCal, MS and the whole ATLAS detector as a function of the \piv mean proper lifetime (c$\tau$) for \mbox{$\mid\eta\mid$~$<2.5$}. The sample with $m_{\rm{H}}=140$\,GeV and $m_{\pi_{\rm{v}}}=20$\,GeV is used.}
	\label{fig:prob}
\end{figure}

\begin{figure}[th!]
	\centering
	\begin{subfigure}[b]{0.48\textwidth}
	\includegraphics[width=\textwidth]{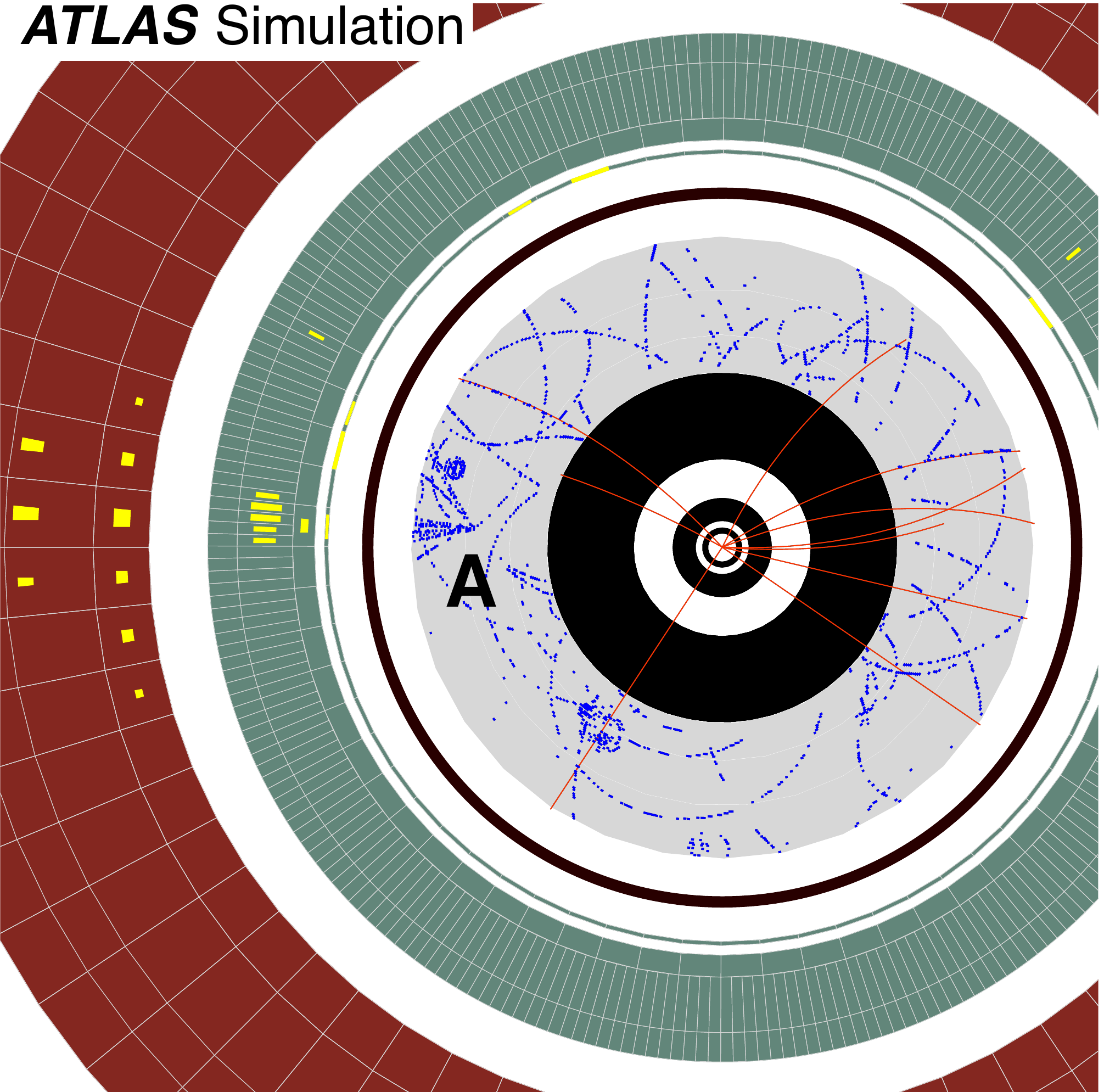}
	\caption{}
	\end{subfigure}
	\begin{subfigure}[b]{0.48\textwidth}
	\includegraphics[width=\textwidth,height=0.995\textwidth]{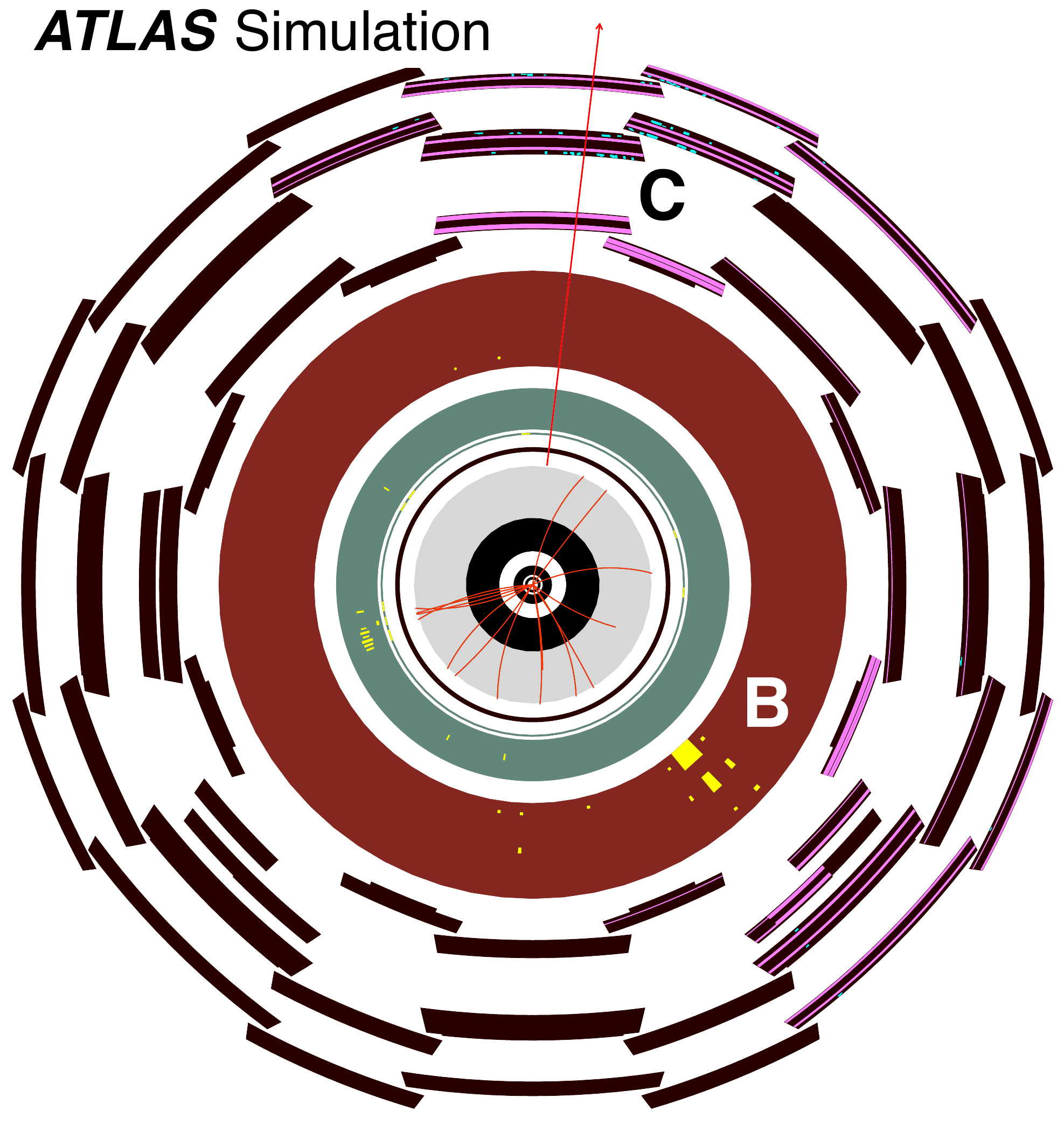}
	\caption{}
	\end{subfigure}
	\caption{Event displays of two simulated $H\to\pi_{\rm{v}}\pi_{\rm{v}}$ events with different \piv decay signatures to hadronic jets. The sample with $m_{\rm{H}}=140$\,GeV and $m_{\pi_{\rm{v}}}=20$\,GeV is used. The ATLAS detectors are depicted in different colours: black for the pixel detector and the SCT, light gray for the TRT, green for the \ECal, red for the \HCal and again black for the MS. The solenoid and the pre-sampler calorimeter are shown as the black and green rings between the ID and the \ECal. 
(a) Event display with one \piv decay in the ID (A). The display focuses on the innermost part of the ATLAS detector. TRT hits clearly indicate the presence of a displaced decay on the left side of the detector, matching well with the energy deposited in the \ECal. 
(b) Event display with one \piv decay in the \HCal (B) with the signature of little activity in the ID and a jet with all of the energy deposit in the \HCal and a second decay in the MS (C). The RPC hits are displayed as sky-blue dots and the MDT hits as pink lines. The red arrow indicates the direction of the missing transverse momentum that correctly points towards the \piv decay in the MS.}
	\label{fig:EventDisplays}
\end{figure}

Figure \ref{fig:prob} shows the probability for a \piv to decay in the barrel and \Endcaps fiducial regions of the ATLAS detector as a function of the \piv mean proper lifetime. There is a substantial probability (greater than 10\%) for decays to occur in the ATLAS detector volume for a wide range of mean proper lifetimes. For example, with a mean proper lifetime of 2\,m the total fraction of \piv decays in the ATLAS detector is 58\%, of which 13\% are in the ID, 12\% are in the \ECal, 13\% are in the \HCal and 20\% are in the MS.

With the exception of the novel triggers described in this paper, the ATLAS triggers were designed to select physics objects originating at (or near) the IP. Events with long-lived particles present many challenges for the trigger system: for example, muons from \piv decays do not have associated tracks in the ID, while jets from \piv decays may have relatively low energy, not have the usual pattern of energy deposition and arrive later than expected. Typical signatures of \piv decays to hadronic jets occurring in the ID, HCal and MS are shown in figures \ref{fig:EventDisplays}(a) and \ref{fig:EventDisplays}(b). 

In order to improve the trigger efficiency, a set of signature-driven triggers were developed to select events based on the detector signature of the displaced hadronic jets. Three regions of the detector are considered: decays beyond the pixel layers to the \ECal, decays in the \HCal and decays from the end of the \HCal to the MS middle station. For each of these regions a unique decay signature exists that can be exploited to select events with displaced decays in the region $|\eta|<2.5$ for which tracking information is available.
\section{Decays in the inner detector and electromagnetic calorimeter}
\begin{figure}[ht!] 
	\centering
	\begin{subfigure}[b]{0.47\textwidth}
	\includegraphics[width=\textwidth]{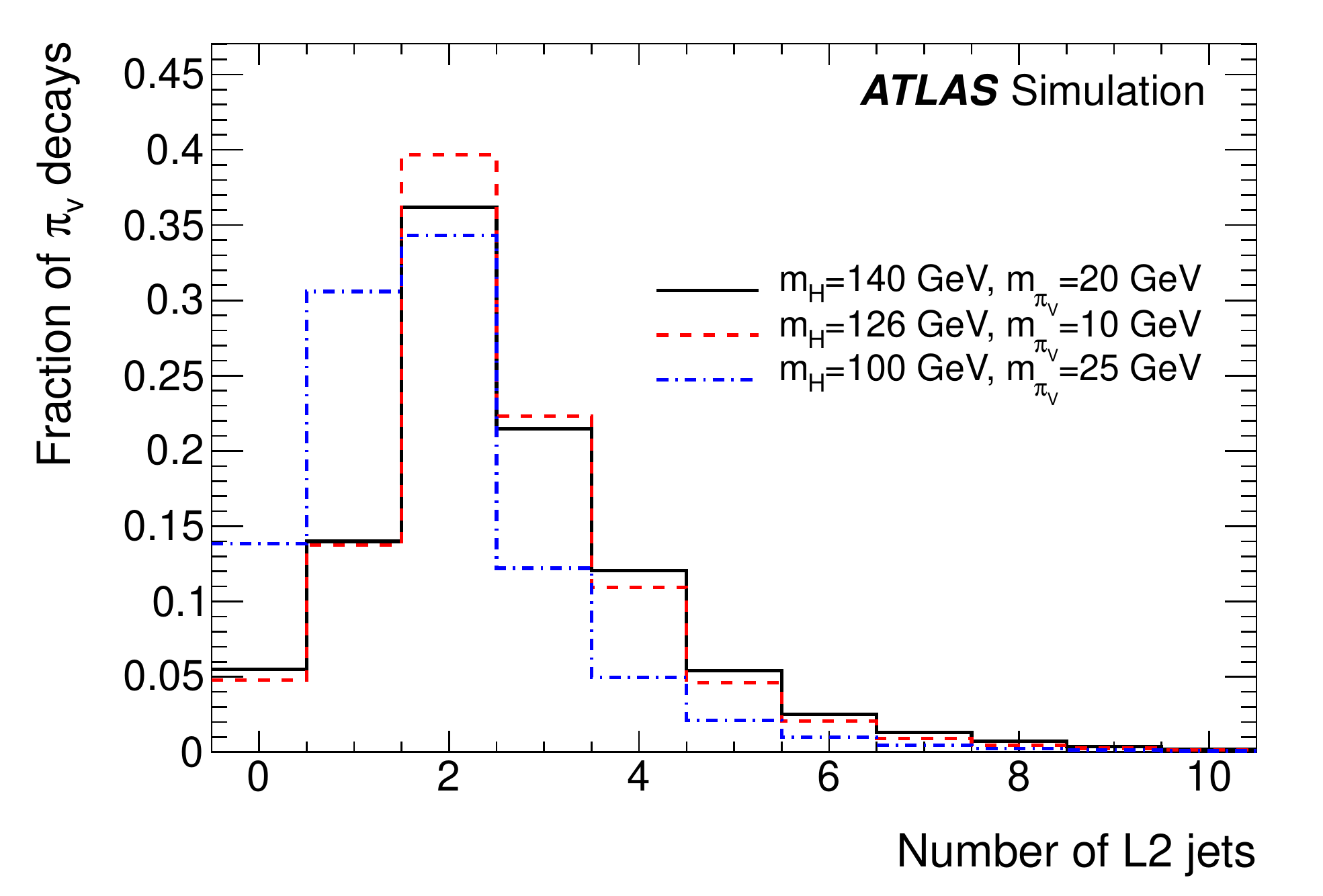}
	\caption{}
	\end{subfigure}
	\begin{subfigure}[b]{0.47\textwidth}
	\includegraphics[width=\textwidth]{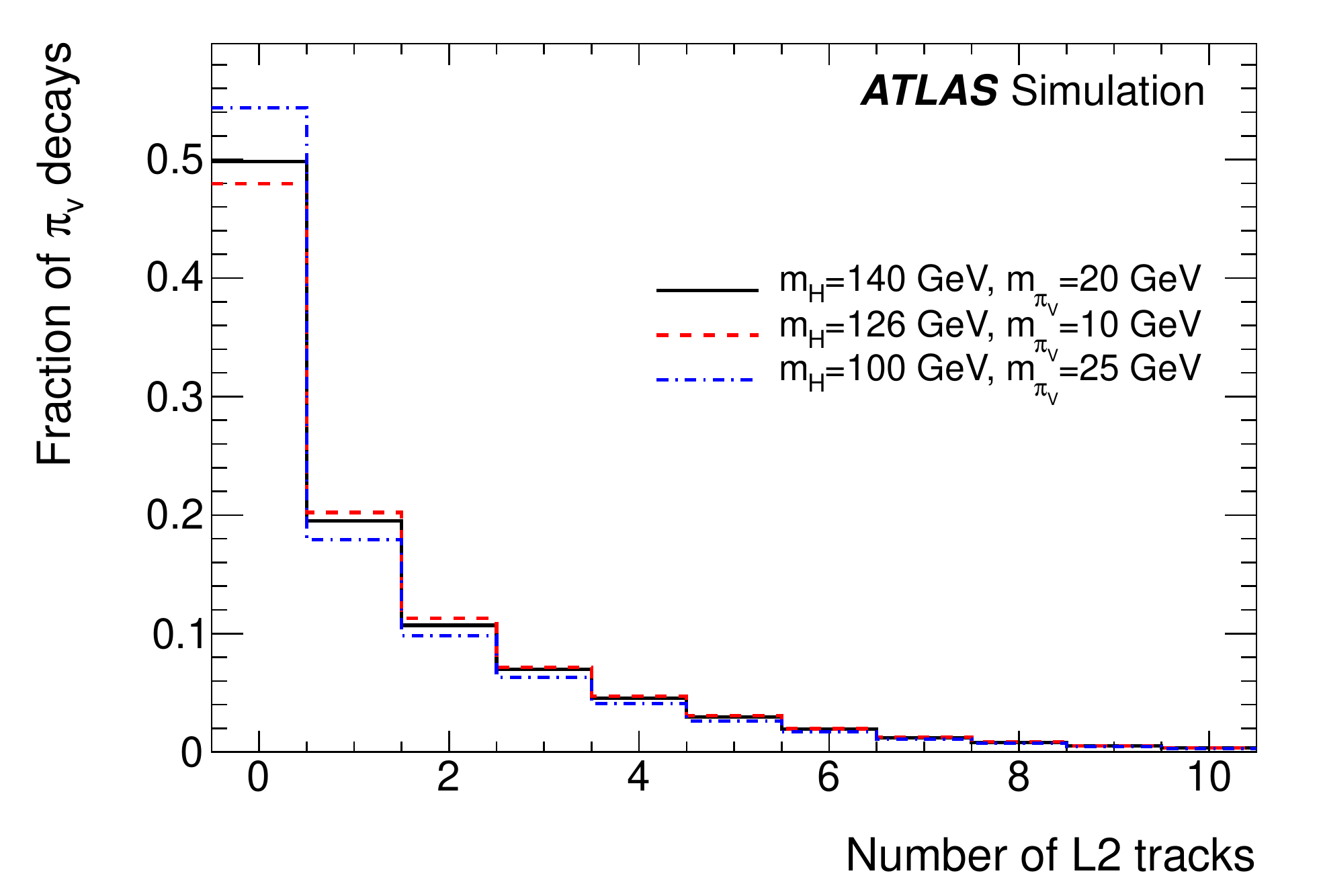}
	\caption{}
	\end{subfigure}
	\caption{(a) Number of reconstructed jets at L2 with $\ET>25$\,GeV and \mbox{$\mid\eta\mid$~$<2.5$} in events where both \piv decays are beyond the pixel detector and before the \HCal volume. (b) Number of reconstructed tracks in the ID at L2 with $\pT>0.8$\,GeV in a $(0.2\times0.2)$ region in ($\Delta\eta\times\Delta\phi$) around the jet axis of a L2 jet with $\ET>25$\,GeV and \mbox{$\mid\eta\mid$~$<2.5$} in events where both \piv decays are beyond the pixel detector and before the \HCal volume.}
	\label{fig:TracklessTriggerPlots}
\end{figure}

A decay topology with two jets characterizes most of the events where both \piv decays are in the \ECal volume, as shown in figure~\ref{fig:TracklessTriggerPlots}(a). Decays that occur beyond the pixel detector produce jets that have no tracks reconstructed by the L2 track reconstruction algorithm. This results in a characteristic signature of a jet isolated with respect to tracks in the ID. A large fraction of jets originating from \piv decays beyond the pixel detector have no reconstructed tracks with $\pT>0.8$\,GeV in a region of $(0.2\times0.2)$ in ($\Delta\eta\times\Delta\phi$) centred on the jet axis, as can be seen in figure~\ref{fig:TracklessTriggerPlots}(b). A jet fulfilling these requirements is defined as a ``trackless jet''.\\

The trackless jet signature provides a powerful handle for identifying jets from displaced decays. Because tracking is not available until the HLT, a different approach is necessary to get a sustainable L1 output rate while not compromising the signal efficiency. In signal events there is a $\sim40$\% probability to produce a muon in the final state. Thus, the multi-jet background can be significantly reduced by requiring the muon to be contained within the jet cone. In multi-jet events produced via SM processes, the muon will typically be prompt and thus leave a track in the ID, while in \piv events the muon is produced close to the \piv decay vertex and can be highly displaced from the IP. 
Requiring both a jet trigger item and a muon trigger item at L1 is therefore a good handle for limiting the L1 output rate. The two objects are required to be in the same detector region at L2 and to be isolated with respect to tracks reconstructed in the ID.
\begin{figure}[t!] 
	\centering
	\includegraphics[width=\textwidth]{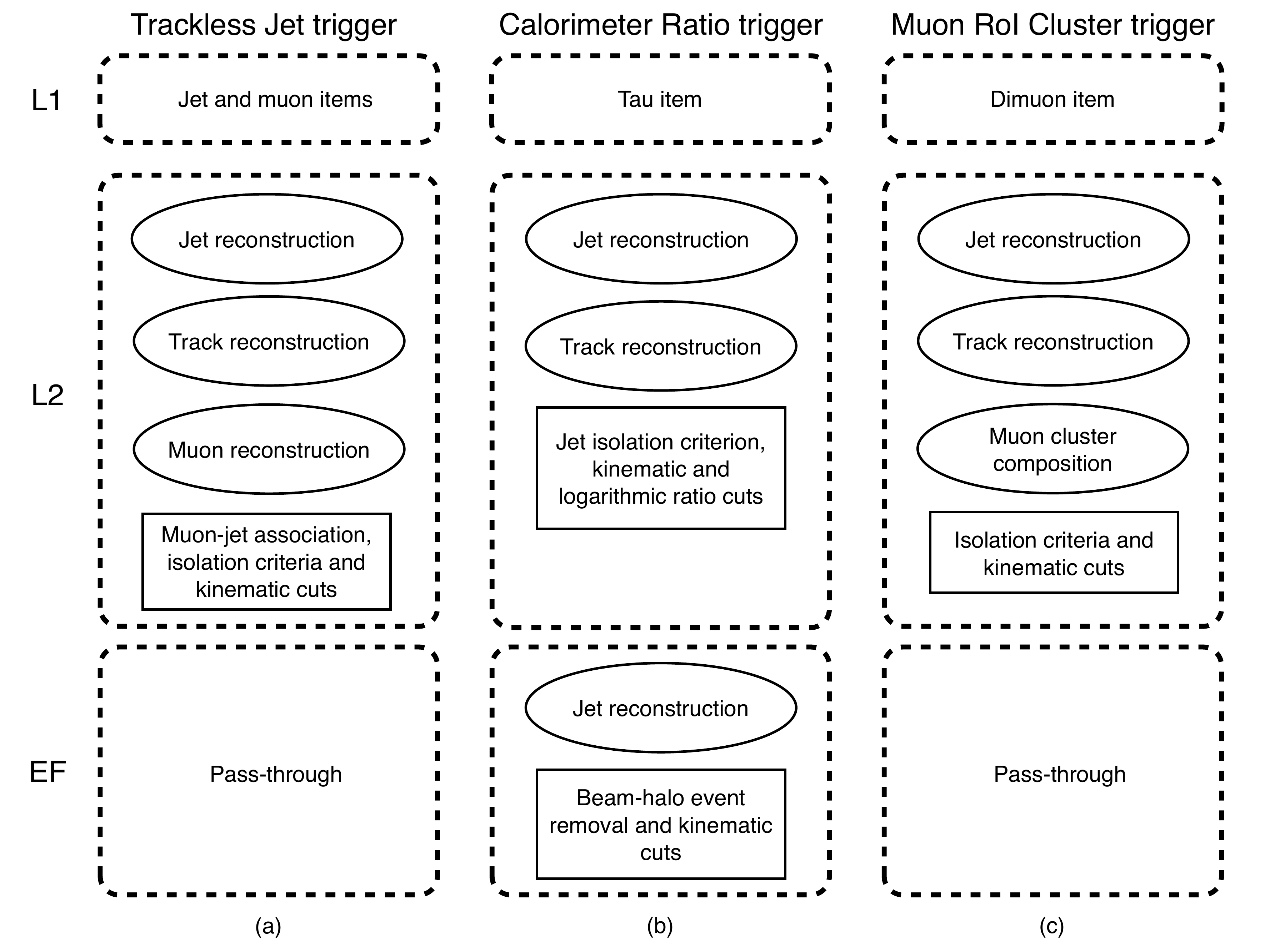}
	\caption{Schematic diagrams of the sequence of trigger algorithms used to select long-lived neutral particles decaying throughout the ATLAS detector volume: Trackless Jet trigger (a), Calorimeter Ratio trigger (b) and Muon \roi Cluster trigger (c). The ovals represent trigger algorithms to reconstruct features while rectangles represent trigger algorithms that apply cuts using the reconstructed features.}
	\label{fig:TriggerDiagrams}
\end{figure}

The Trackless Jet trigger implementation requires a muon with $\pT>10$\,GeV and a jet with $\ET>20$\,GeV at L1. The L2 jet and muon reconstruction algorithms are then employed in the corresponding \rois to confirm and refine the physics objects identified at L1. Because the particles from displaced decays do not have reconstructable tracks in the ID, the L2 muon reconstruction algorithm is restricted to run only in the MS. 
The criteria at L2 are $\pT>10$\,GeV for the muon and $\ET>25$\,GeV for the jet, where the calorimeter response is calibrated at the electromagnetic (EM) scale, which is tuned to correctly measure the energy of EM showers. A combinatorial algorithm performs a geometrical matching in a $\Delta R\equiv\sqrt{(\Delta\eta)^{2}+(\Delta\phi)^{2}}$~=~0.4 region between the muon and jet objects. The final criterion of no tracks in the ID with $\pT>0.8$\,GeV in a $(0.2\times0.2)$ region in ($\Delta\eta\times\Delta\phi$) around the jet direction is made at L2. No further selection is applied at the EF. A schematic diagram of the sequence of algorithms employed in the Trackless Jet trigger is depicted in figure~\ref{fig:TriggerDiagrams}(a).

\section{Decays in the hadronic calorimeter}
Decays of neutral particles in the outer layers of the \ECal or in the \HCal result in little or no energy deposited in the \ECal (figure \ref{fig:EventDisplays}) and consequently in an anomalously large value of the hadronic to electromagnetic energy ratio, $\CaloRatio$, of the jet. 
Since the neutral \piv has travelled through the tracking volume without interacting before decaying in the calorimeter, no reconstructed tracks connecting the jet to the primary IP are expected. 
Another feature of this topology is that the pair of hadronic jets doesn't have time to spatially separate before depositing its energy in the \HCal and this results in a very narrow single jet being reconstructed.
The ``Calorimeter Ratio trigger'' is designed to select this signature.

\begin{figure}[h!]
	\centering
	\begin{subfigure}[b]{0.48\textwidth}
	\includegraphics[width=\textwidth]{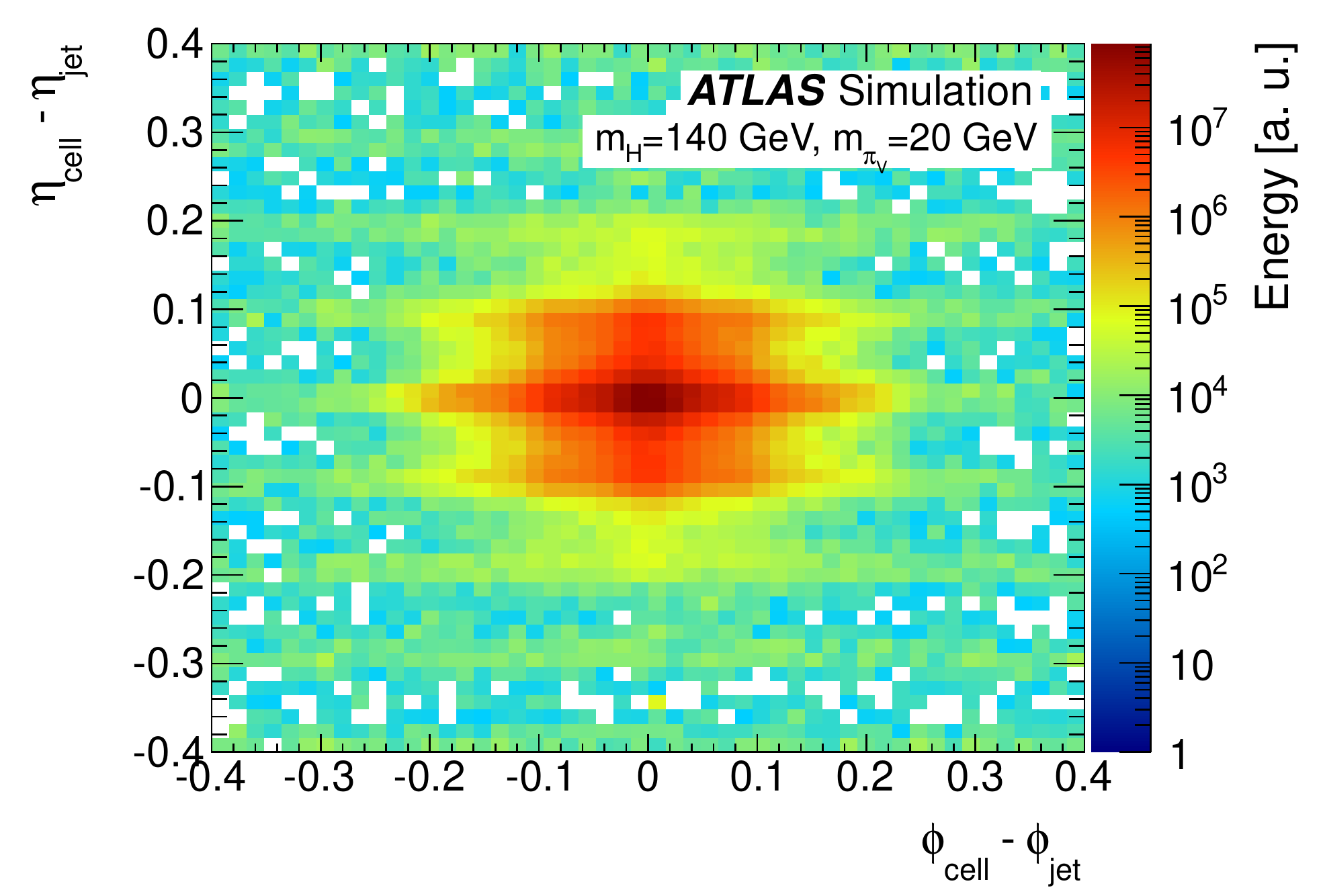}
	\caption{}
	\end{subfigure}
	\begin{subfigure}[b]{0.48\textwidth}
	\includegraphics[width=\textwidth]{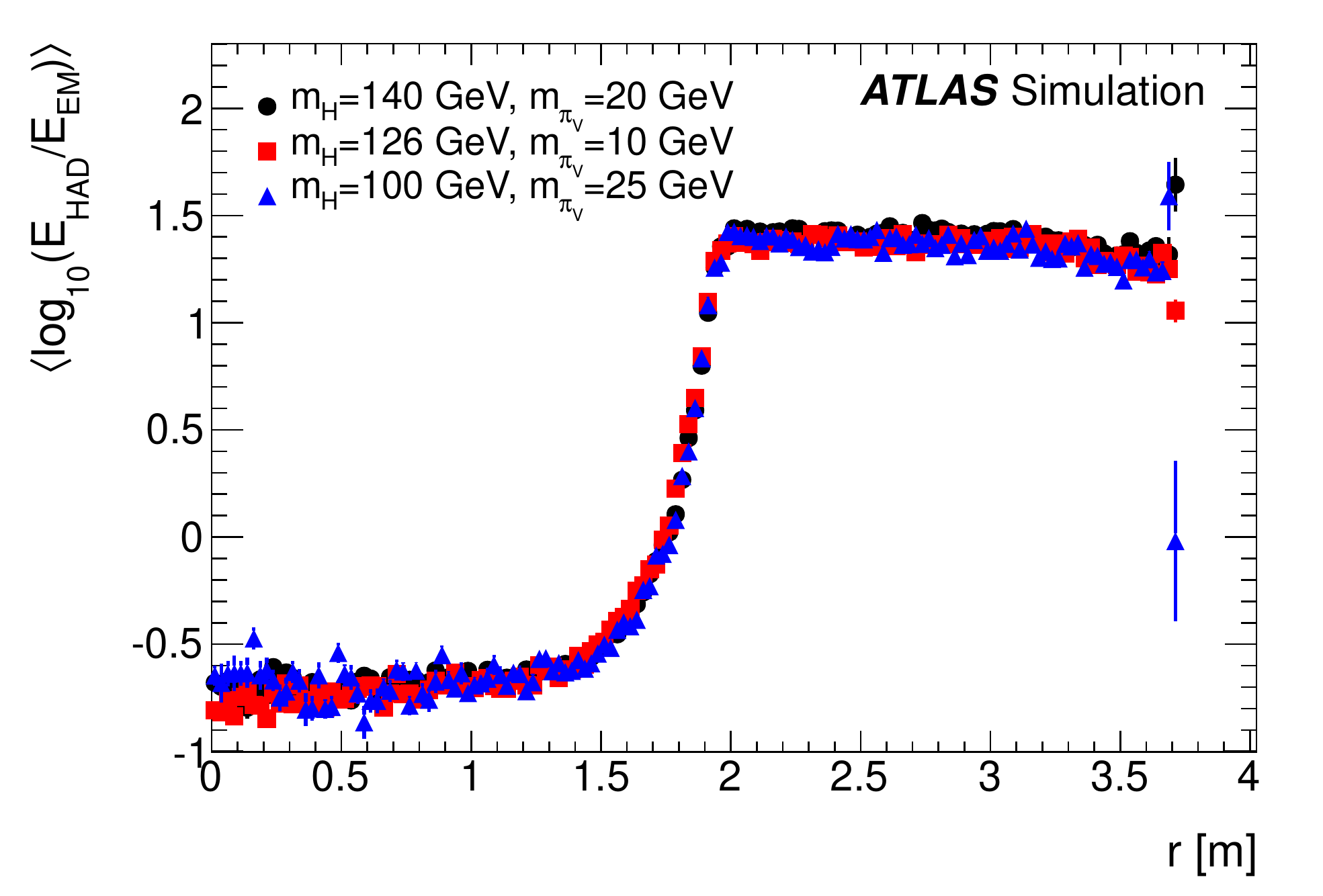}	
	\caption{}
	\end{subfigure}
	\caption{(a) Energy deposition in the calorimeter cells around the jet axis (0,0) for \piv decays in the \HCal. (b) Mean value of $\LogCaloRatio$ for all L2 jets in the barrel calorimeter coming from \piv decays as a function of the \piv radial decay distance, $r$.}
	\label{fig:CaloRatio_Calo}
\end{figure}

\begin{figure}[t!]
	\centering
	\begin{subfigure}[b]{0.48\textwidth}
	\includegraphics[width=\textwidth]{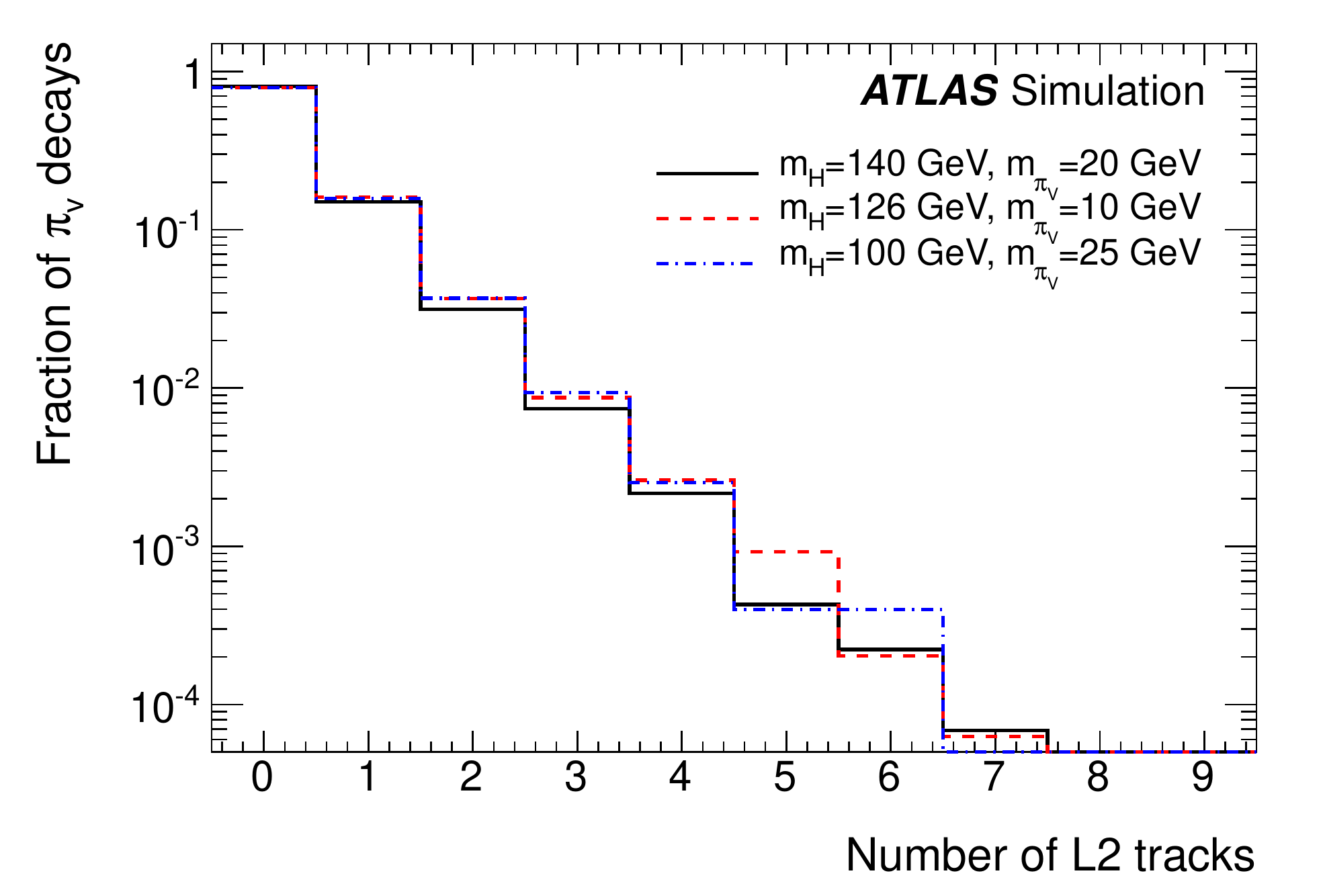}
	\caption{}
	\end{subfigure}
	\begin{subfigure}[b]{0.48\textwidth}
	\includegraphics[width=\textwidth]{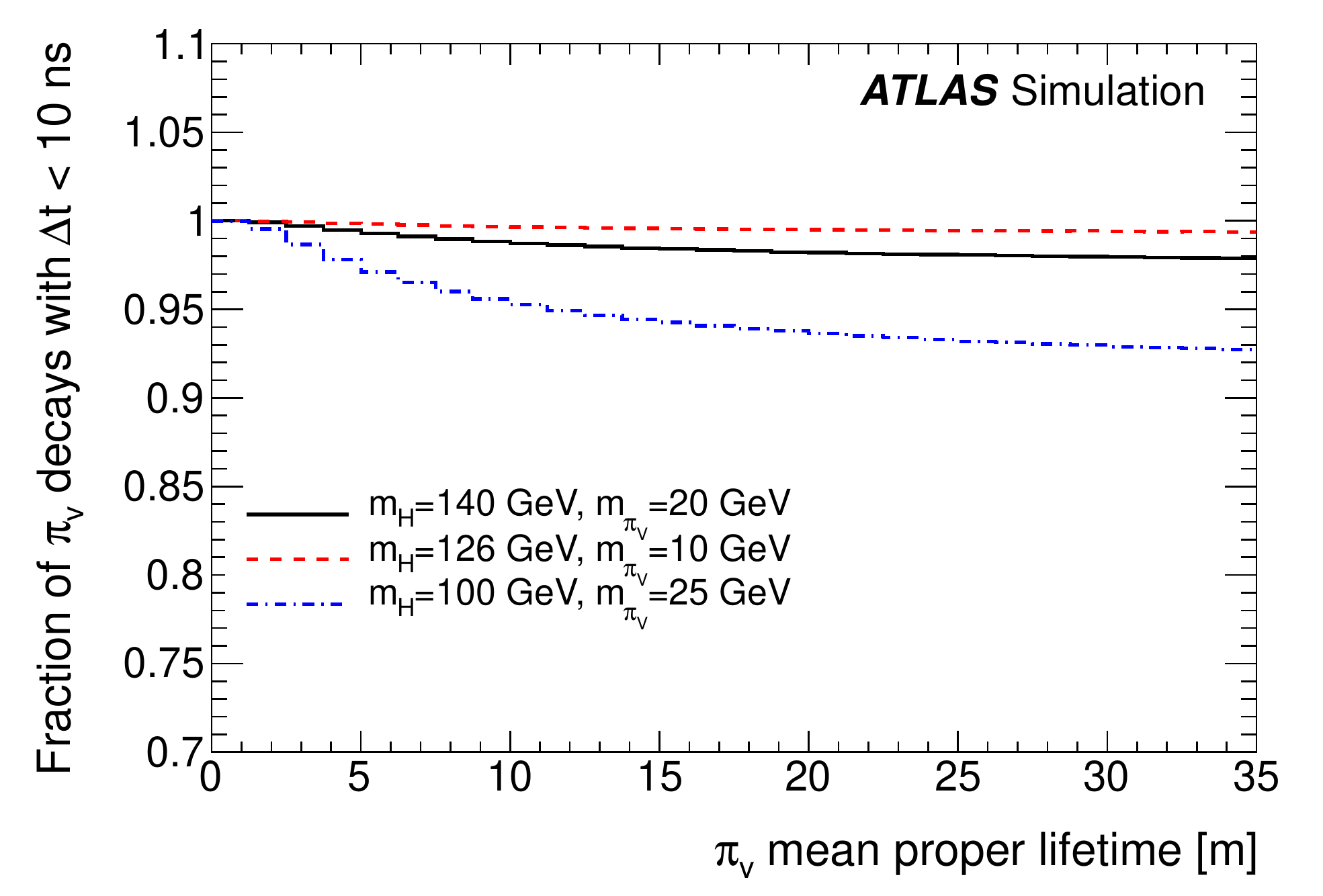}
	\caption{}
	\end{subfigure}
	\caption{(a) Distribution of the number of reconstructed tracks in the ID at L2 with $\pT>1$\,GeV in a ($0.2\times0.2$) region in ($\Delta\eta\times\Delta\phi$) around the jet axis of a L2 jet with $\LogCaloRatio>1.2$ and \mbox{$\mid\eta\mid$~$<1.5$} in events where the \piv decays in the \HCal. (b) Fraction of \piv decays in the \HCal with a delay of $\Delta t<10$\,ns as a function of the \piv mean proper lifetime.}
	\label{fig:CaloRatio_TrackAndTiming}
\end{figure}

As can been seen in figure~\ref{fig:CaloRatio_Calo}(a), almost all of the energy from jets originating from \piv decays in the \HCal is contained within a cone of $\Delta R=0.1$ around the jet axis. For this reason, a L1 $\tau$-lepton trigger is preferable to a L1 jet trigger because the latter selects events in a ($\Delta\eta\times\Delta\phi$) region of (0.4$\times$0.4) while the former selects, by design, jets with energy contained in a region of (0.2$\times$0.2), which better matches the narrow jets produced when a \piv decays in the \HCal. The $\tau$-lepton triggers also have a lower energy threshold than the jet triggers, resulting in an increase of the signal efficiency relative to the multi-jet background acceptance. Figure~\ref{fig:CaloRatio_Calo}(b) shows the mean value of the $\LogCaloRatio$ distribution for all L2 jets from \piv decays in the barrel calorimeter as a function of the \piv radial decay distance. As the \piv decays close to the end of the \ECal, the logarithmic ratio changes from a characteristic negative value to a positive value. The average value of $\LogCaloRatio$ plateaus at $\sim$\,1.5 for radii between $\sim$\,1.9\,m and $\sim$\,3.6\,m, at which point the \piv decay occurs too close to the outer radius of the \HCal to reconstruct a jet. The Gaussian width ($\sigma$) of the $\LogCaloRatio$ distribution beyond 2\,m is 0.35. Because jets with $\LogCaloRatio>1.2$ are produced by \piv decays in the calorimeter, the ID should not contain any tracks pointing to a signal jet. The result of applying the L2 tracking algorithm is that about 80\% of the jets with $\LogCaloRatio>1.2$ have no tracks with $\pT>1$\,GeV reconstructed in a region of (0.2$\times$0.2) in ($\Delta\eta\times\Delta\phi$) around the jet axis, as shown in figure~\ref{fig:CaloRatio_TrackAndTiming}(a).
 
The \piv particles from Higgs decay have $\beta < 1$ and arrive in the calorimeter later than a $\beta=1$ particle. The signal peak amplitude is thus shifted later in time, and this affects the raw energy measurement. The \online Optimal Filtering Algorithm \cite{OFA} determines this time shift and provides a corrected energy at L2 that is within 1\% of the true energy for time shifts of up to 12\,ns \cite {HCalSignalReconstruction}. Figure \ref{fig:CaloRatio_TrackAndTiming}(b) from the ATLAS simulation of the calorimeter trigger, which incorporates the Optimal Filtering Algorithm and energy correction algorithms, shows that for the $m_{\rm{H}}$ and $m_{\pi_{\rm{v}}}$ combinations considered, more than 93\% of the \piv decays in the HCal have a time shift of less than 10\,ns for a broad range of mean proper lifetimes.

Beam-halo muons that undergo bremsstrahlung as they traverse the \HCal are a potential source of background for the Calorimeter Ratio trigger. 
If an energetic bremsstrahlung radiation is emitted by the muon in the \HCal, the L1 trigger logic interprets this energy deposit as a jet and the event is passed to L2; no energy is found in the \ECal and therefore the event is accepted at L2. In order to discard these events, an additional algorithm, exploiting the non-collision-like timing signature in the \HCal, is implemented in the EF of the Calorimeter Ratio trigger. If at least three \HCal cells with $E>240$\,MeV are found with a difference in timing of 5\,ns with respect to a particle travelling at $\beta=1$, the event is rejected. Only  cells within $|\Delta\phi|<0.2$ and $\Delta R>0.3$ from the leading jet axis are considered. The removal of beam-halo muons reduces the Calorimeter Ratio trigger rate by almost 50\% without compromising the signal acceptance.

The Calorimeter Ratio trigger is built by requiring a $\tau$-lepton object with $\ET>40$\,GeV at L1. The threshold at 40\,GeV corresponds to the lowest unprescaled L1 $\tau$-lepton item used during data taking in 2012. The L2 jet and track reconstruction algorithms are then employed using the L1 $\tau$-lepton \roi as input. The L2 reconstructed jet is required to have $\ET>30$\,GeV, with the calorimeter response calibrated at the EM scale. Two additional criteria are applied: that there be no tracks with $\pT>1$\,GeV reconstructed in a $(0.2\times0.2)$ region in ($\Delta\eta\times\Delta\phi$) around the jet direction and that the jet $\LogCaloRatio>1.2$. At the EF, the standard anti-$k_t$ jet-finding algorithm \cite{AntiKt} is added to refine the jet object properties and $\ET>35$\,GeV is required. The final criterion removes beam-halo events. A schematic diagram of the sequence of algorithms employed in the Calorimeter Ratio trigger is depicted in figure~\ref{fig:TriggerDiagrams}(b).

\section{Decays in the muon spectrometer}
Decays occurring at the outermost regions of the \HCal or in the MS result in a large number of charged hadrons traversing a narrow region of the MS. These events are characterized by a cluster of L1 muon \rois centred around the \piv line of flight.
\begin{figure}[t!]
	\centering
	\begin{subfigure}[b]{0.47\textwidth}
	\includegraphics[width=\textwidth]{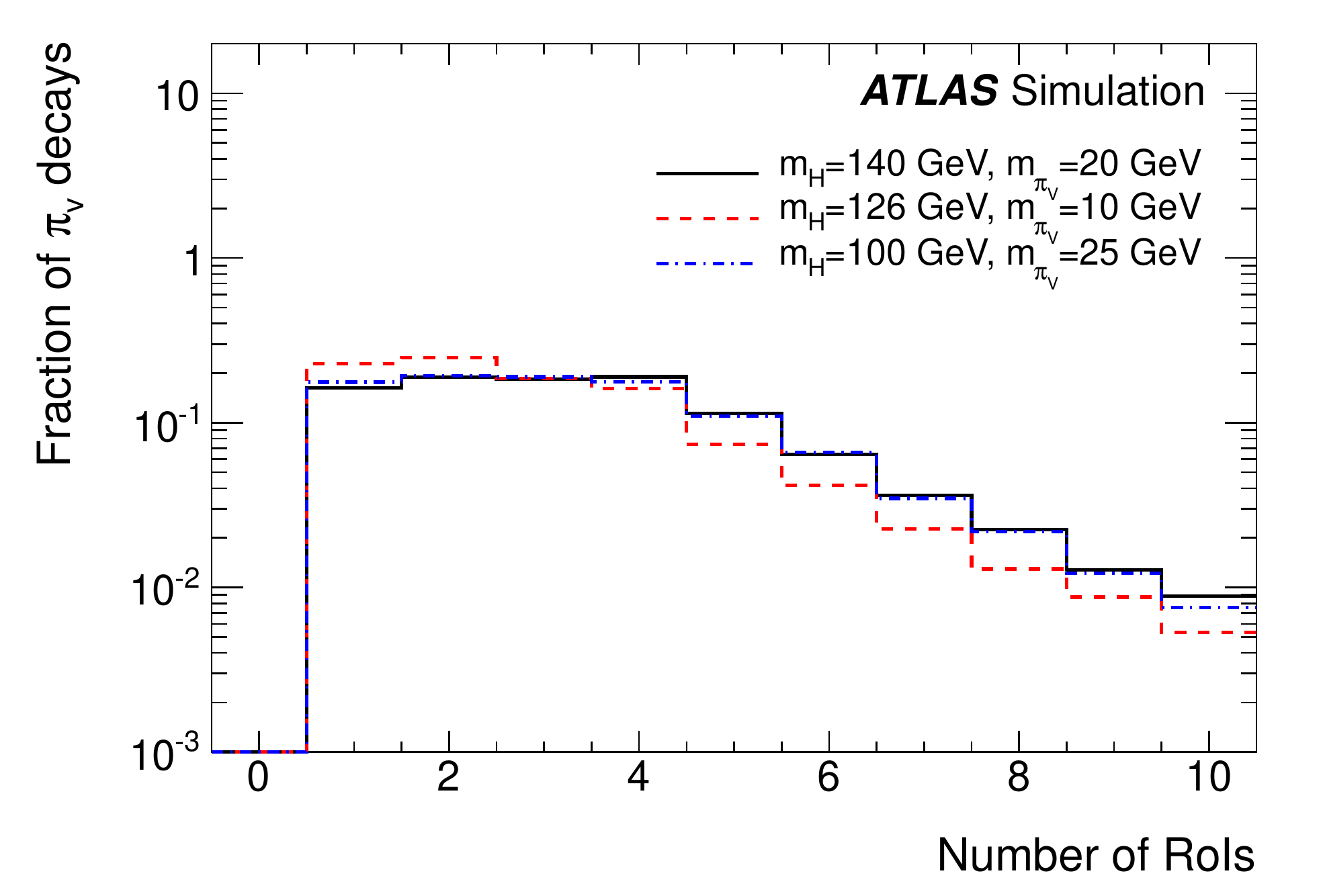}
	\caption{}
	\end{subfigure}
	\begin{subfigure}[b]{0.47\textwidth}
	\includegraphics[width=\textwidth]{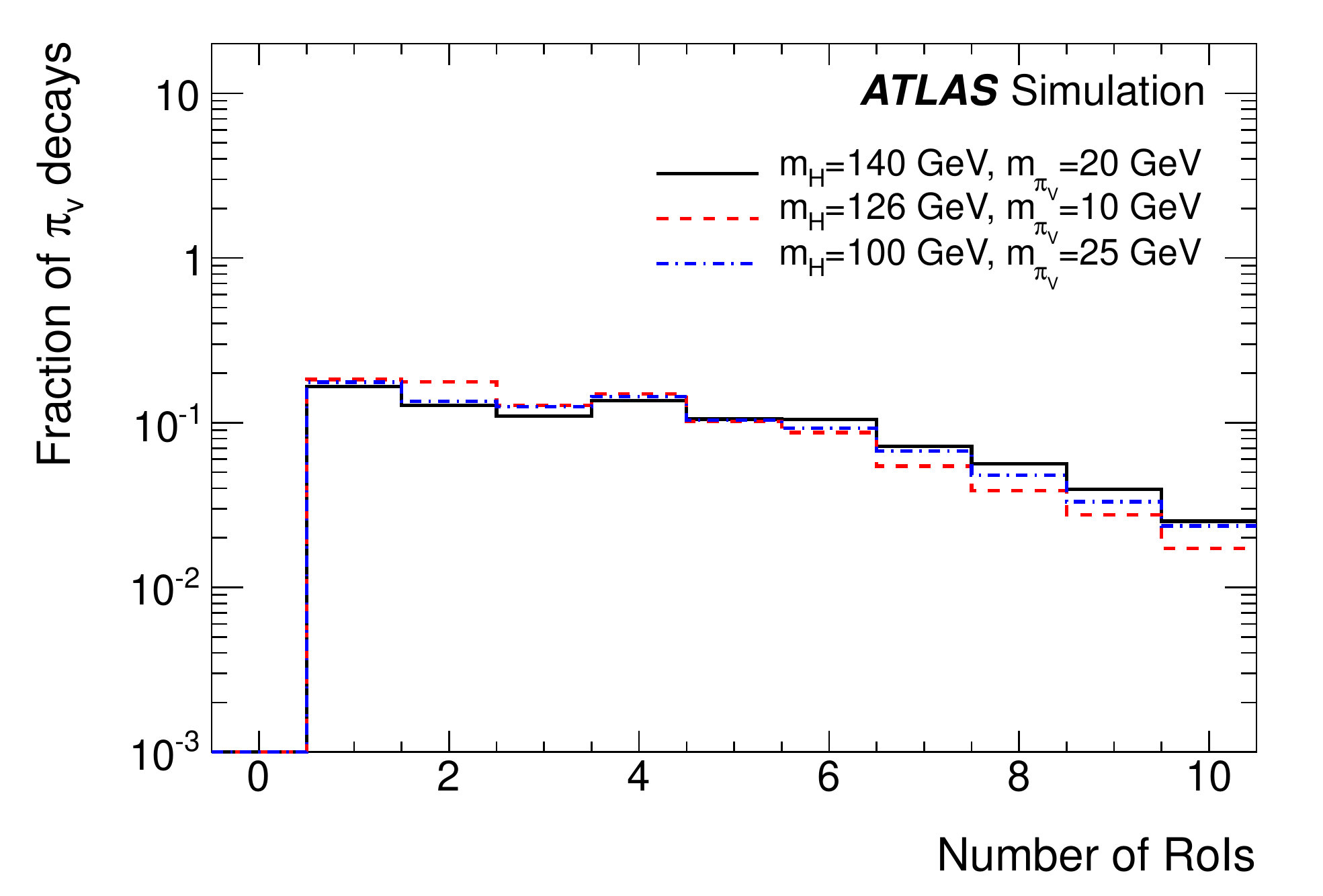}
	\caption{}
	\end{subfigure}
	\caption{Distribution of the total number of \rois found in events with a \piv decaying in (a) the MS barrel and (b) the MS \Endcaps.}
	\label{fig:nRoI}
\end{figure}
\begin{figure}[t!]
	\centering
	\begin{subfigure}[b]{0.47\textwidth}
	\includegraphics[width=\textwidth]{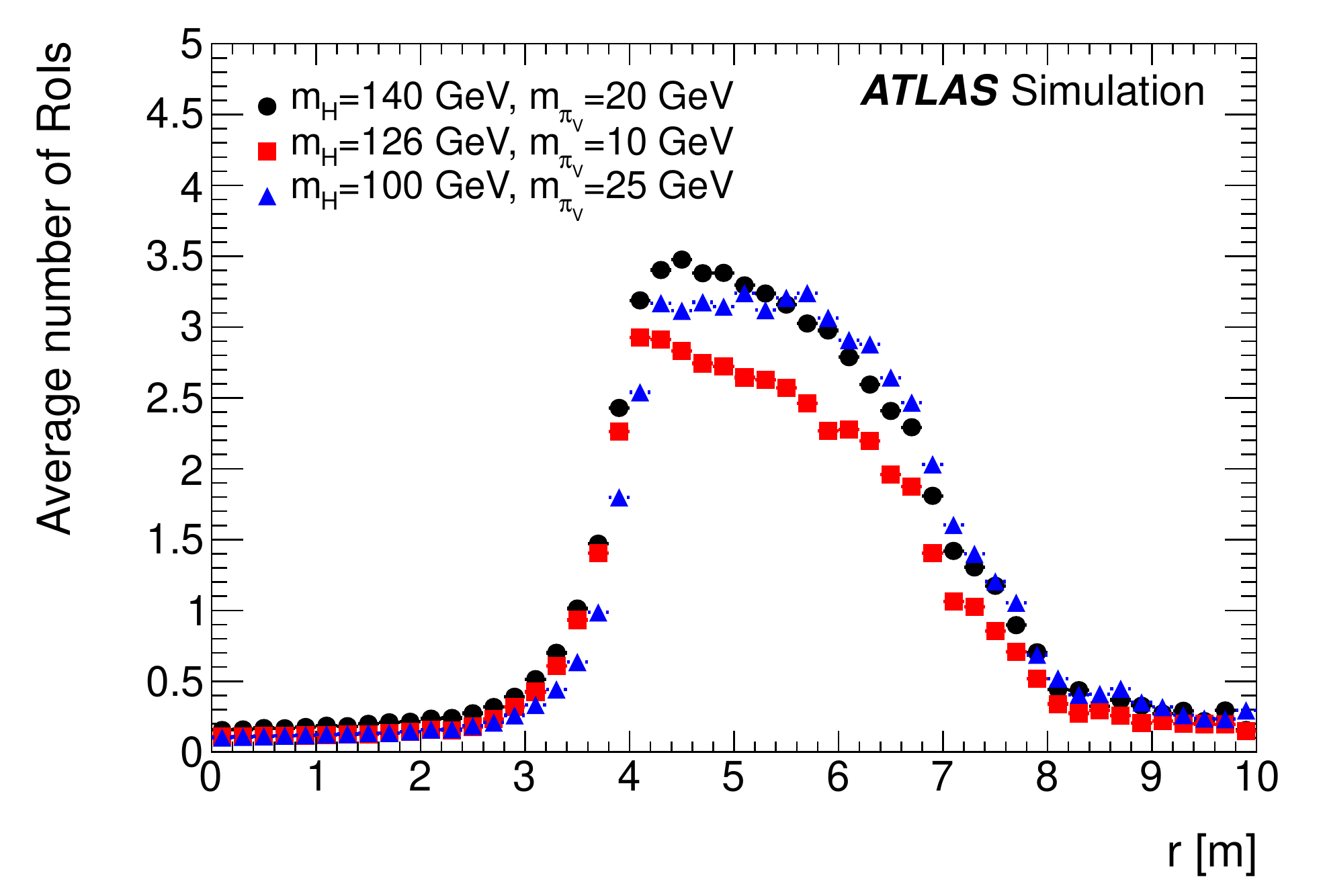}
	\caption{}
	\end{subfigure}
	\begin{subfigure}[b]{0.47\textwidth}
	\includegraphics[width=\textwidth]{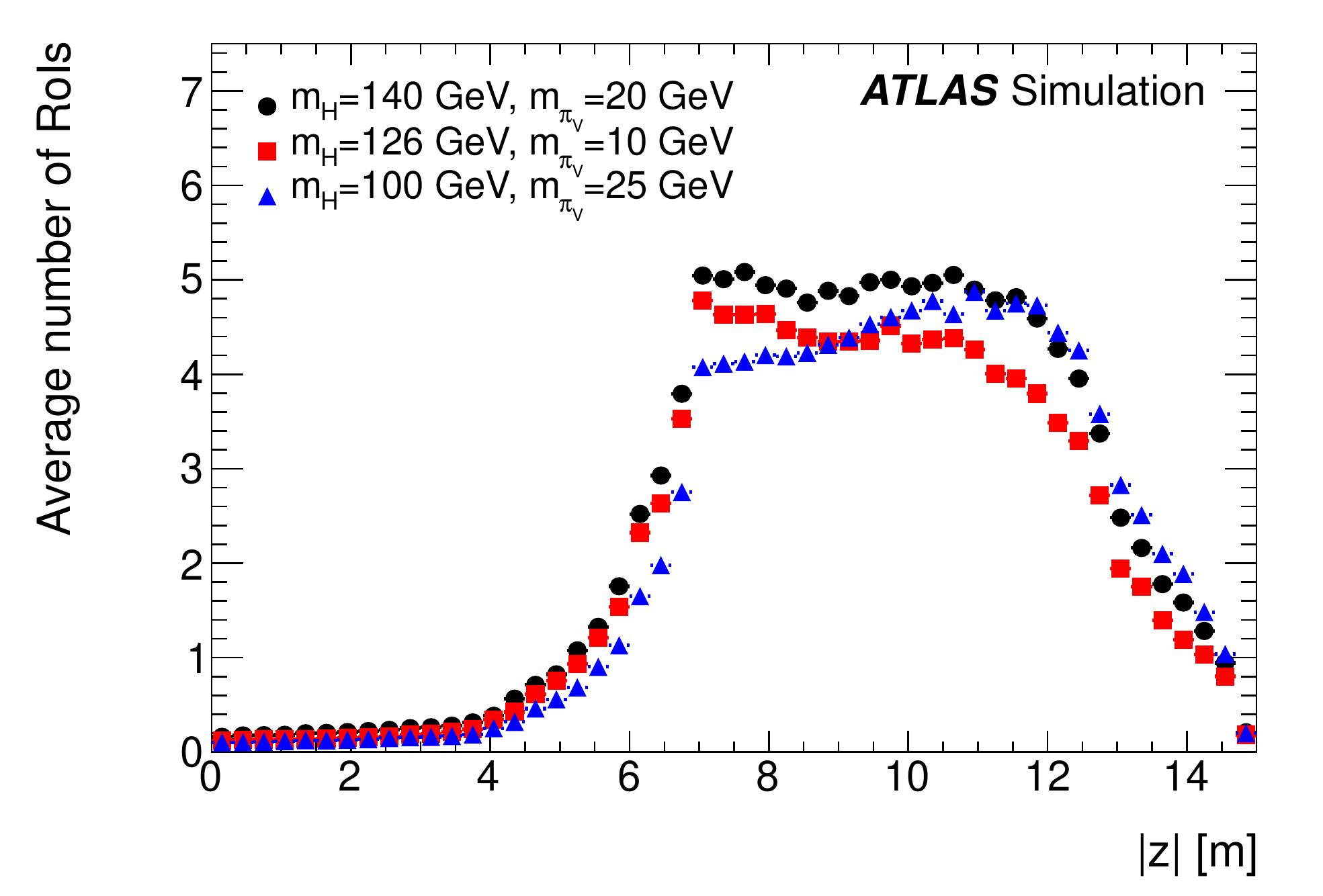}
	\caption{}
	\end{subfigure}
	\caption{Average number of L1 muon \rois in a cone of $\Delta R=0.4$ around the \piv line of flight as a function of  (a) the \piv radial decay position, $r$, in the MS barrel and (b) the $|z|$ coordinate of the \piv decay in the MS \Endcaps.}
	\label{fig:nRoI2}
\end{figure}

Figures~\ref{fig:nRoI}(a) and \ref{fig:nRoI}(b) show the total number of muon \rois found in events with a \piv decaying in the MS barrel or \Endcaps. Larger \roi multiplicities are found in the \Endcaps because the muon \rois have a spatial extent of 0.1 in $\Delta\eta$ and $\Delta\phi$ in the \Endcaps compared to 0.2 in the barrel. Figures~\ref{fig:nRoI2}(a) and \ref{fig:nRoI2}(b) show the average number of L1 muon \rois contained in a cone of $\Delta R=0.4$ around the \piv line of flight as a function of the \piv radial decay distance in the MS barrel and as a function of the $|z|$ decay coordinate in the MS \Endcaps. Figure~\ref{fig:nRoI2} illustrates that a \piv decaying in the MS results in a mean value of approximately three \rois and from four to five \rois clustered around the \piv line of flight  in the barrel and in the \Endcaps, respectively. The mean number of \rois in the MS barrel is less than three \rois for the sample with $m_{\rm{H}}=126$\,GeV and $m_{\pi_{\rm{v}}}=10$\,GeV due to the higher \piv boost, which leads to narrower jets in the MS. For \piv decays close to the end of the \HCal ($r\sim4$\,m in the barrel and $|z|\sim6$\,m in the \Endcaps), the average number of muon \rois contained in the cone increases rapidly due to the charged-particle tracks from the \piv decay passing through the calorimeter and entering the MS. Once the \piv decays occur in the MS, the number of \rois remains approximately constant until the \piv decays are close to the muon trigger plane ($r\sim7$\,m in the barrel and $|z|\sim13$\,m in the \Endcaps), at which point the charged hadrons are not spatially separated enough to give multiple, unique \rois. Moreover, decays occurring beyond the first trigger plane do not give \rois because hits are required in both the first and second trigger planes.

A cluster of muon \rois is defined as a $\Delta R=0.4$ radius region in the MS barrel (\Endcaps) containing at least three (four) muon \rois. This characteristic signature is exploited at the HLT to select events with \piv decays at the outermost regions of the \HCal or in the MS. The SM backgrounds from punch-through jets and bremsstrahlung from muons can be suppressed by requiring the \roi cluster to be isolated with respect to both calorimeter jets and tracks in the ID.

\begin{figure}[ht!]
	\centering
	\begin{subfigure}[b]{0.48\textwidth}
	\includegraphics[width=\textwidth]{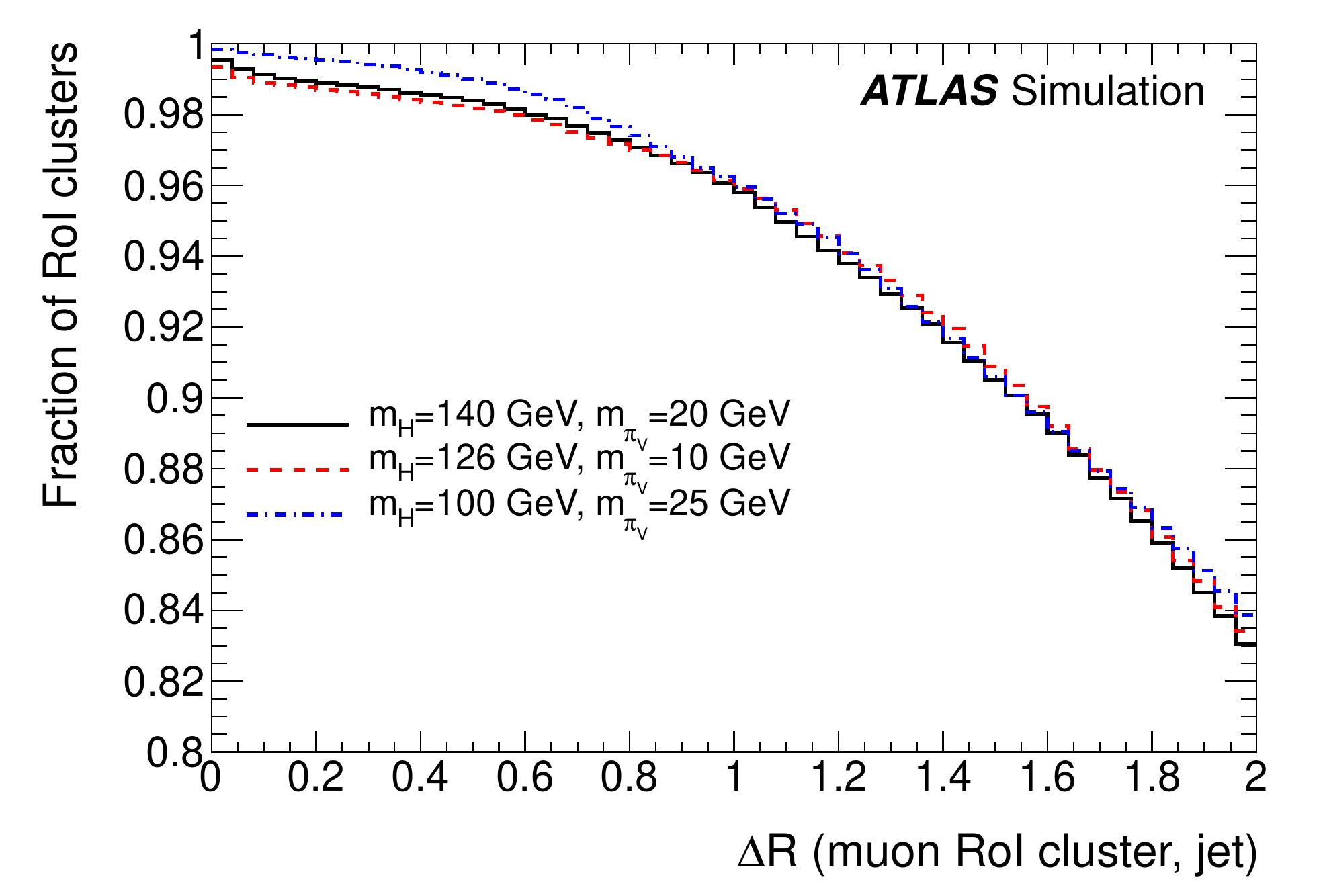}
	\caption{}
	\label{fig:RoIClusterA}
	\end{subfigure}
	\begin{subfigure}[b]{0.48\textwidth}
	\includegraphics[width=\textwidth]{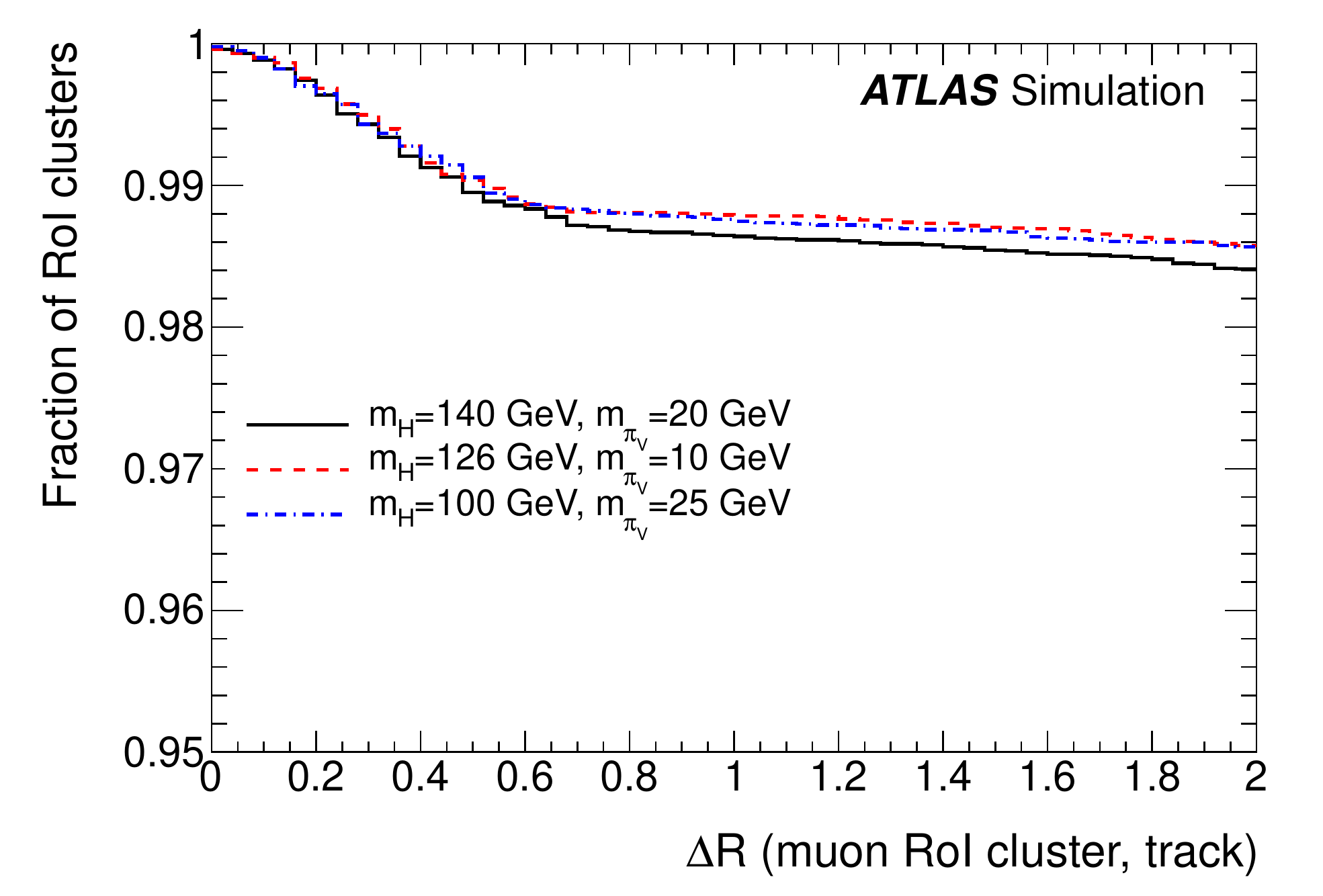}
	\caption{}
	\label{fig:RoIClusterB}
	\end{subfigure}
	\caption{Fraction of \roi clusters accepted in signal events as a function of $\Delta R$ between (a) the centre of the \roi cluster and the jet direction for L2 jets with $\LogCaloRatio<0.5$ and (b) the centre of the \roi cluster and the track direction for L2 tracks with $\pT>5$\,GeV.}
	\label{fig:MuonCluster_isolation}
\end{figure}

Figure~\ref{fig:MuonCluster_isolation}(a) shows the fraction of events accepted as a function of the $\Delta R$ distance from the nearest jet to the \roi cluster centre in signal events. The acceptance is relatively flat up to values of $\Delta R\sim0.7$. Requiring the muon \roi cluster to be isolated from jets within $\Delta R<0.7$ results in a signal loss of less than $\sim3$\%. It should be noted that the \roi cluster is only required to be isolated from jets that have $\LogCaloRatio<0.5$ in order to increase the signal efficiency for events in which the \piv decays at the end of the \HCal, depositing energy and thus producing both a jet with significant hadronic energy and a muon \roi cluster.

Figure~\ref{fig:MuonCluster_isolation}(b) shows the fraction of events accepted at L2 as a function of $\Delta R$ between the centre of the \roi cluster and the nearest track in the ID with $\pT>5$\,GeV. Because the \piv particles are both long lived, high-$\pT$ tracks are not expected to be reconstructed in the ID. The acceptance is found to be approximately constant across a wide range of $\Delta R$ values. Requiring isolation within a region of $\Delta R<0.4$ results in an acceptance loss of less than $\sim1$\%.

The timing of the L1 muon trigger is centred on the arrival time of the bunch crossing in ATLAS, so that the trigger is active for 12.5\,ns before and 12.5\,ns after the bunch crossing. To ensure that the Muon \roi Cluster trigger is associated with the correct bunch crossing, the critical parameter to examine is the time delay between \piv decays in the MS and a particle produced in the same interaction travelling with $\beta=1$. The \piv decay products have $\beta\simeq1$. Thus, in determining the time delay between the \piv arrival and a $\beta=1$ particle, only the distance the \piv travels from the IP to the decay point needs to be considered.

The efficiency of the RPC trigger was measured, as a function of the time shift $\Delta t$, in a test beam~\cite{H8TB}. The RPC time resolution has also been measured for the 2011 data and found to be in good agreement with the test beam results. The L1 muon trigger accurately matches events to their bunch crossings for time delays of less than approximately 6\,ns. Time delays beyond 6\,ns result in some of the triggers being associated with the next bunch crossing in a predictable Gaussian manner. If the trigger signal from the \piv is not associated with the correct bunch crossing, the event is lost.

The expected \online acceptance of the L1 trigger in the barrel MS for the three benchmark mass points can be calculated\footnote{The distances to the first muon trigger plane are used for the calculation of the online quantities: $L = 7$\,m and $L = 11$\,m for $\eta=0$ and $|\eta|=1$, respectively.} from the measured bunch-crossing identification efficiency, given in ref.~\cite{H8TB}, and the $\beta$ distribution of the \piv. The results are shown in table~\ref{tab:bceff} for \piv particles emitted at $\eta=0$ and $|\eta|=1$. Similar results are obtained for the trigger system in the \Endcaps. The table also gives the trigger acceptances obtained from the simulation, which are systematically shifted to larger values because in the barrel trigger simulation at L1 there is a time shift of plus 3.125\,ns (one least count) with respect to the \online trigger. As a consequence, the simulation has a correctable, systematically higher acceptance for late trigger signals as shown in table~\ref{tab:bceff} for the three benchmark samples. Analyses using this trigger will need to calculate the corrected efficiency.
\begin{figure}[t!]
	\centering
	\includegraphics[width=0.6\textwidth]{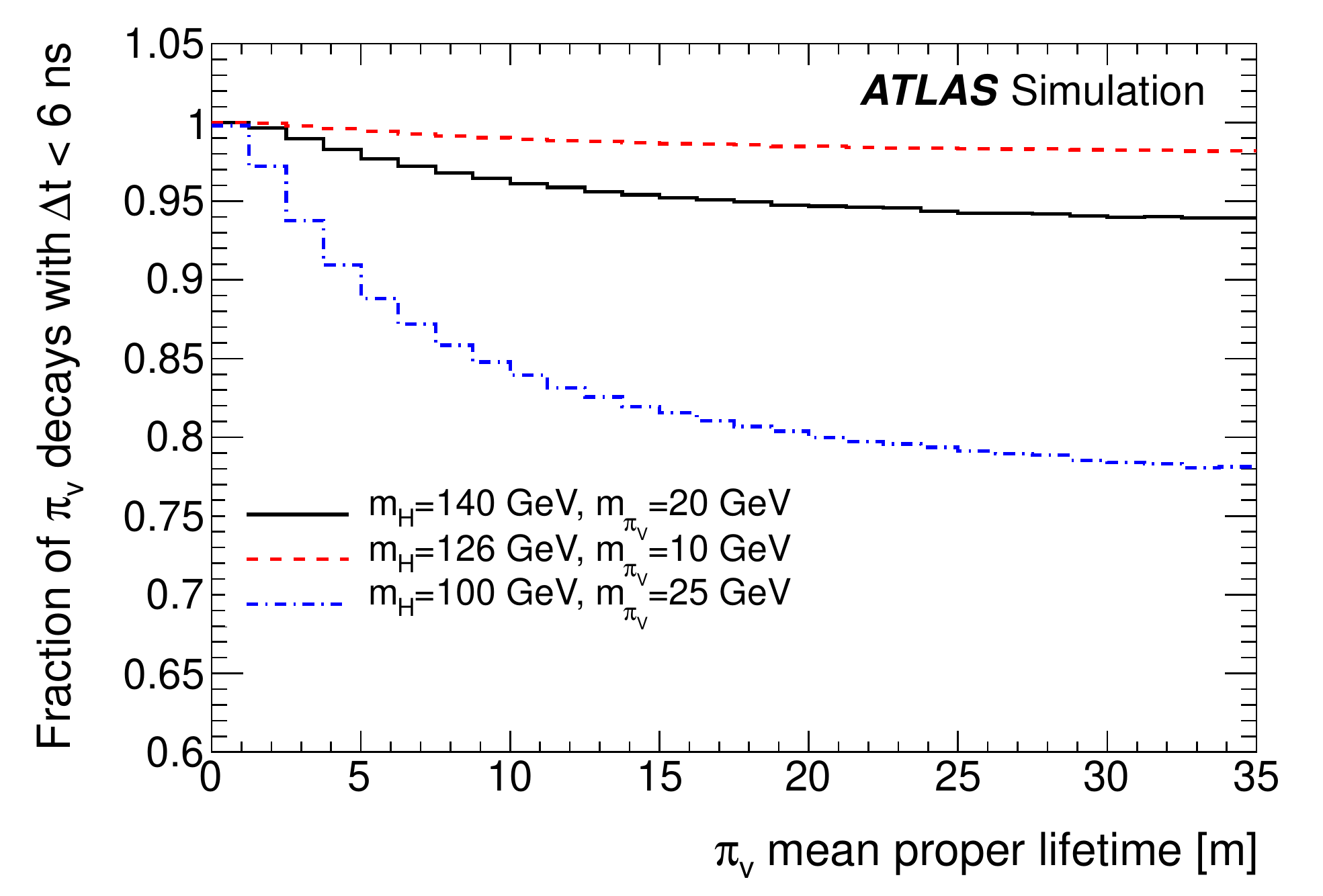}
	\caption{Fraction of \piv decays producing a Muon \roi Cluster trigger associated with the correct bunch crossing as a function of the \piv mean proper lifetime.}
	\label{fig:RoICluster_Timing}
\end{figure}
\begin{table}[ht!]\centering
\begin{tabular}{ccccc}
\hline
Sample & data eff. $\eta=0$  & MC eff. $\eta= 0$ & data eff. $|\eta|=1$  & MC eff. $|\eta|=1$ \\\hline
$m_{\rm{H}}/m_{\pi_{\rm{v}}}=100/25$ GeV & 94.4\% &97.6\% & 88.9\% & 94.9\%  \\\hline
$m_{\rm{H}}/m_{\pi_{\rm{v}}}=126/10$ GeV & 99.8\% &99.9\%& 99.6\% & 99.8\% \\\hline
$m_{\rm{H}}/m_{\pi_{\rm{v}}}=140/20$ GeV & 98.9\% &99.4\%& 97.9\%& 99.0\%  \\\hline
\end{tabular}
\caption{The probability of associating \piv particles emitted at $\eta=0$ and $|\eta|=1$ with the correct bunch crossing for both simulation and data for the three benchmark samples. The statistical uncertainties are negligible.}
\label{tab:bceff}
\end{table}

Figure \ref{fig:RoICluster_Timing} shows the fraction of \piv decays in the MS with a $\Delta t$ of less than 6\,ns as a function of the \piv mean proper lifetime. These are decays that could produce a trigger associated with the correct bunch crossing. The \piv is associated with the correct bunch crossing with an efficiency greater than 75\% for the $m_{\rm{H}}=100$\,GeV and $m_{\pi_{\rm{v}}}=25$\,GeV sample and with an efficiency greater than 94\% for the other mass points. When the mean proper decay length is small compared to the detector dimensions, only very boosted \piv particles have non-negligible probability to decay in the MS. As the mean proper decay length increases, those \piv particles with a smaller boost begin to have some probability to decay in the MS. When the mean proper decay length becomes comparable to the ATLAS detector dimensions all the \piv particles have roughly equal probability to decay in the MS. 

The Muon RoI Cluster trigger is built by requiring two muons with $\pT>10$\,GeV at L1. A cluster of muon \rois is required at L2 for all the events passing the L1 selection. Then, the L2 jet and track reconstruction algorithms are employed using all the jet \rois found in the same event. The jet isolation criterion is then applied, requiring that there be no jets within $\Delta R$ = 0.7 around the muon \roi cluster centre with  $\ET>30$\,GeV  and $\LogCaloRatio<0.5$, where the calorimeter response is calibrated at the EM scale. The track isolation criterion is also applied, requiring no tracks in the ID with $\pT>5$\,GeV within $\Delta R$ = 0.4 of the muon \roi cluster centre. No further selection is applied at the EF. A schematic diagram of the sequence of algorithms employed in the Muon \roi Cluster trigger is depicted in figure~\ref{fig:TriggerDiagrams}(c).

\section{Trigger efficiency on simulated events}
\begin{figure}[th!]
	\centering
	\begin{subfigure}[b]{0.47\textwidth}
	\includegraphics[width=\textwidth]{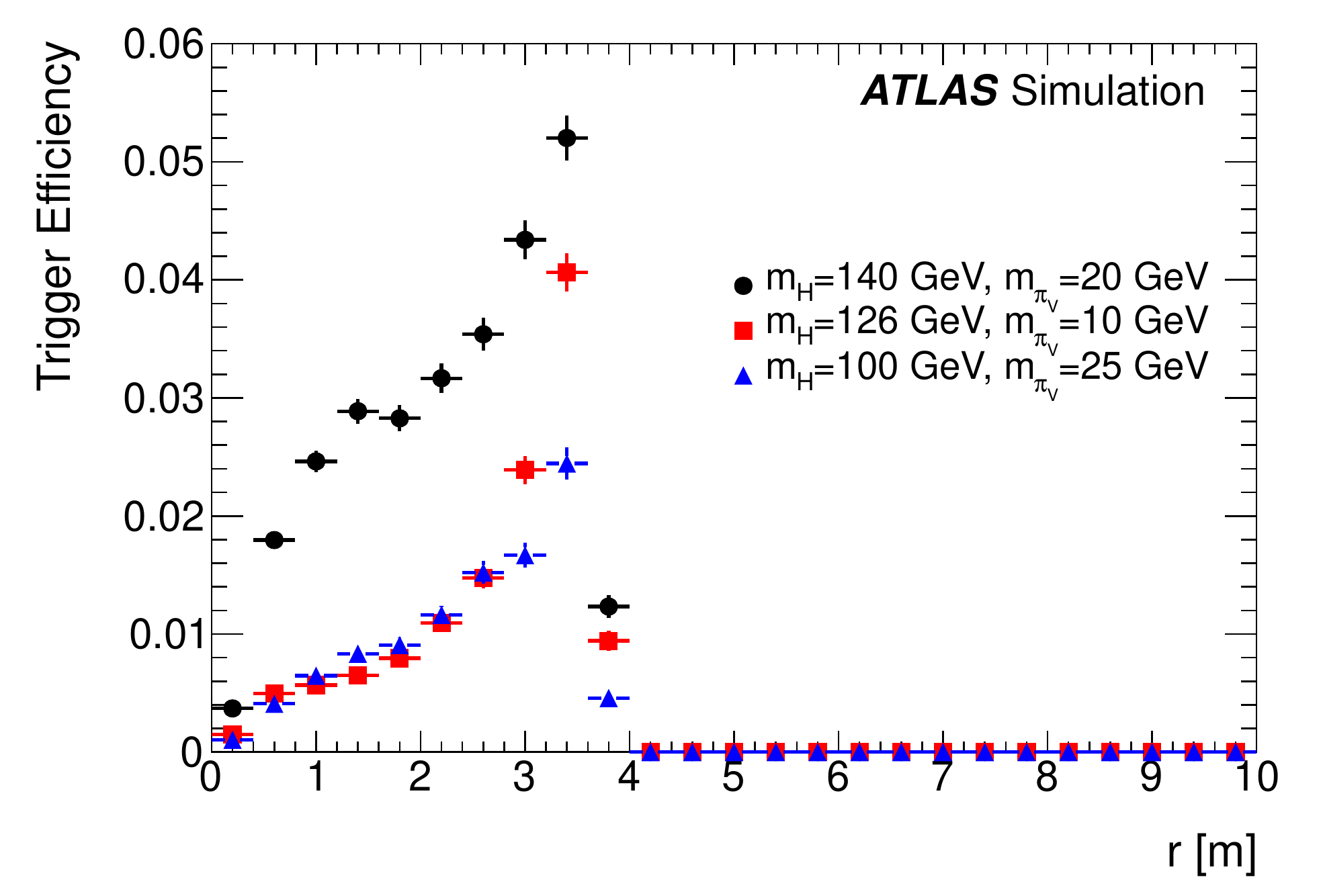}
	\caption{}
	\end{subfigure}
	\begin{subfigure}[b]{0.47\textwidth}
	\includegraphics[width=\textwidth]{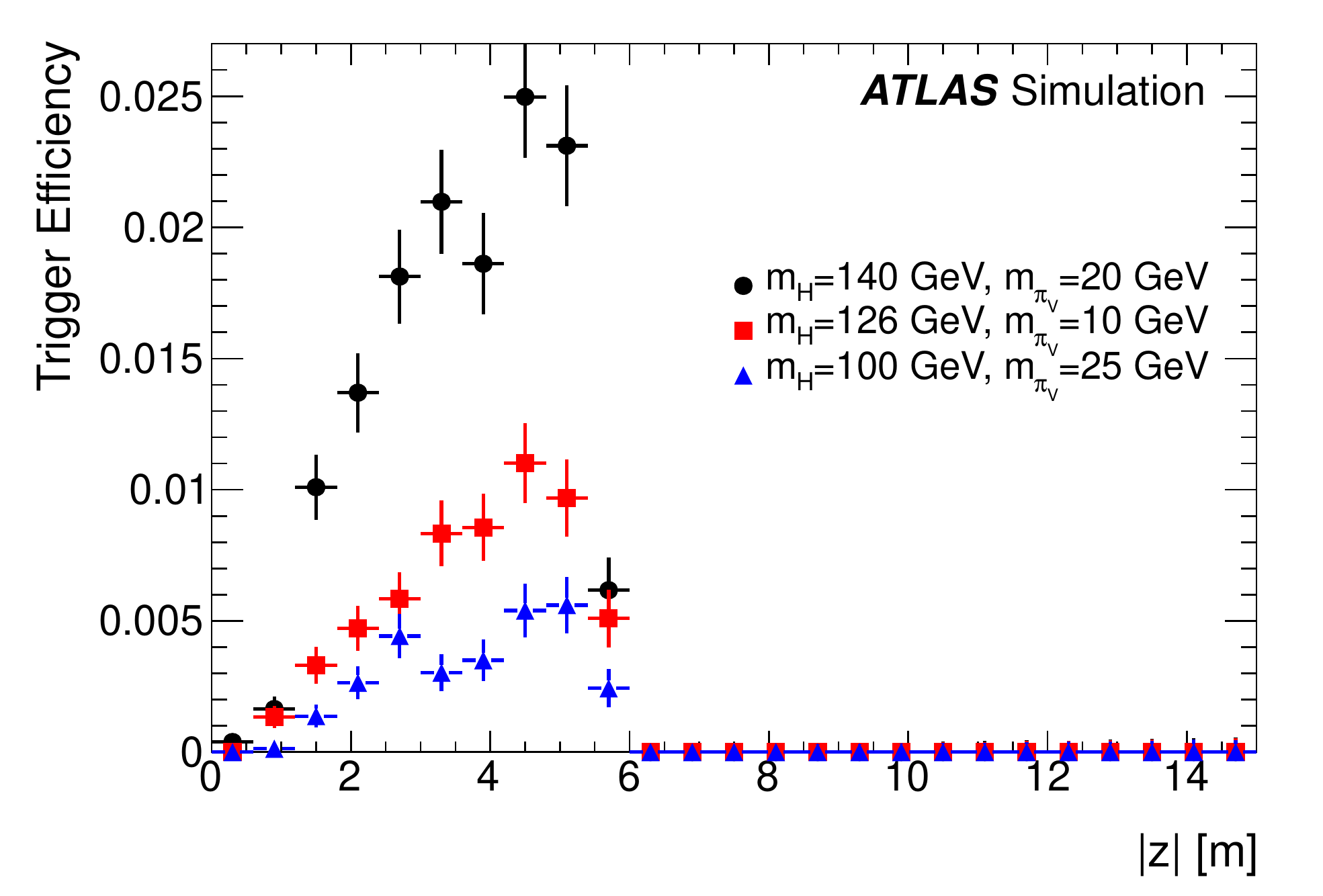}
	\caption{}
	\end{subfigure}
	\caption{Efficiency for the Trackless Jet trigger as a function of (a) the radial decay position, $r$, for \piv decays in the barrel and (b) the $|z|$ position for \piv decays in the \Endcaps.}
	\label{fig:TracklessTriggerEff}
	\vspace{0.4cm}
	\begin{subfigure}[b]{0.47\textwidth}
	\includegraphics[width=\textwidth]{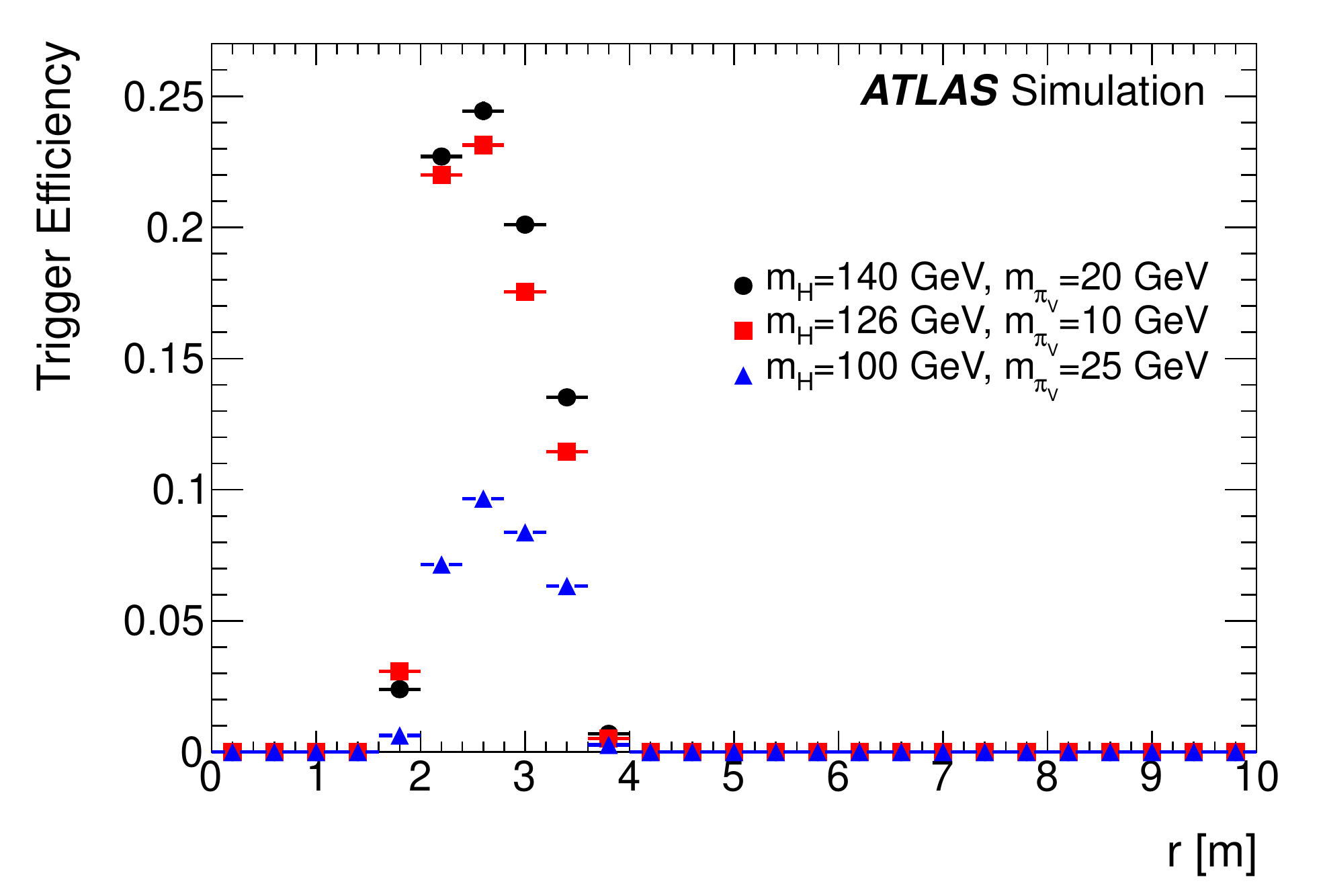}
	\caption{}
	\end{subfigure}
	\begin{subfigure}[b]{0.47\textwidth}
	\includegraphics[width=\textwidth]{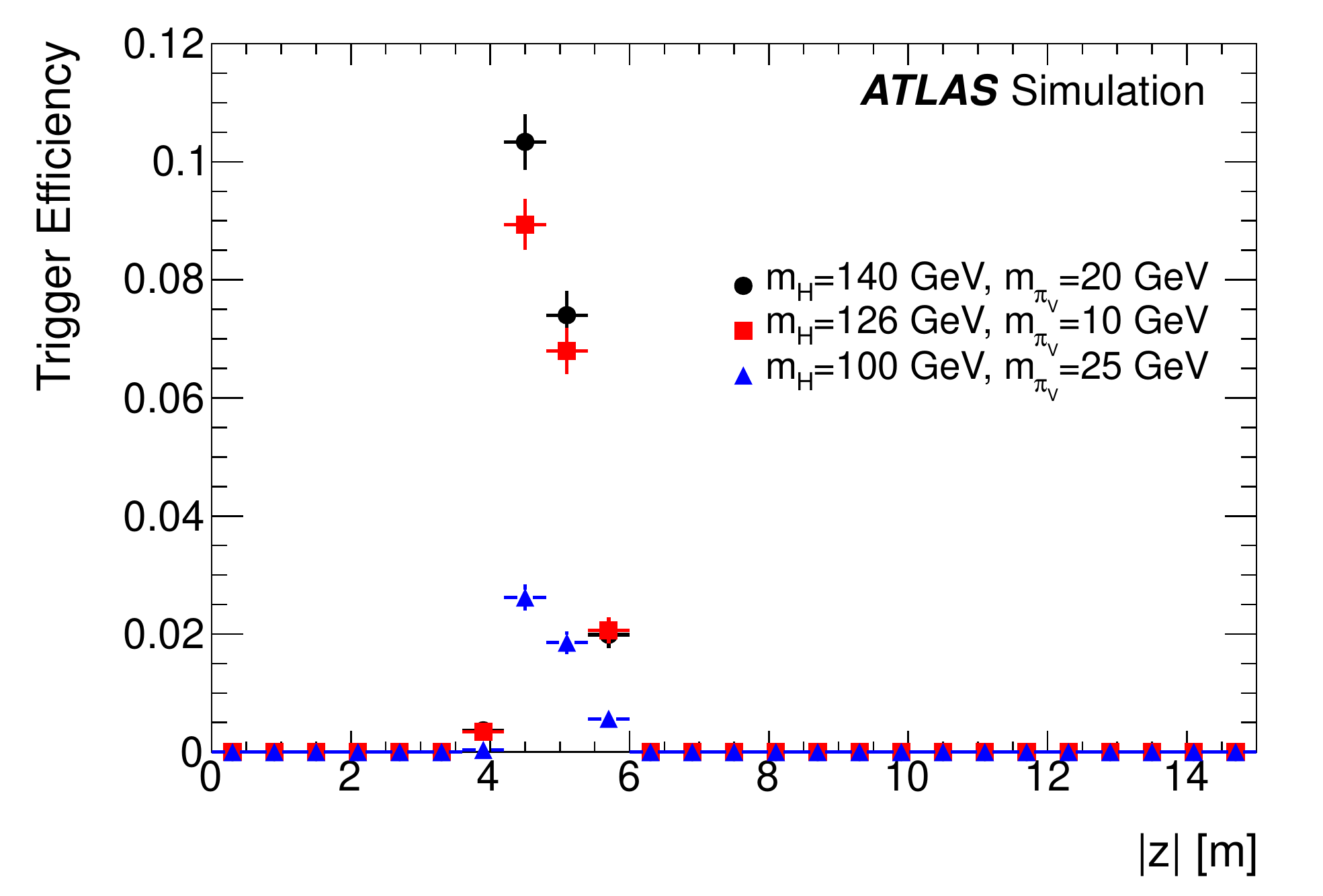}
	\caption{}
	\end{subfigure}
	\caption{Efficiency for the Calorimeter Ratio trigger as a function of (a) the radial decay position, $r$, for \piv decays in the barrel and (b) the $|z|$ position for \piv decays in the \Endcaps.}
	\label{fig:CaloRatioTriggerEff}
	\vspace{0.4cm}
	\begin{subfigure}[b]{0.47\textwidth}
	\includegraphics[width=\textwidth]{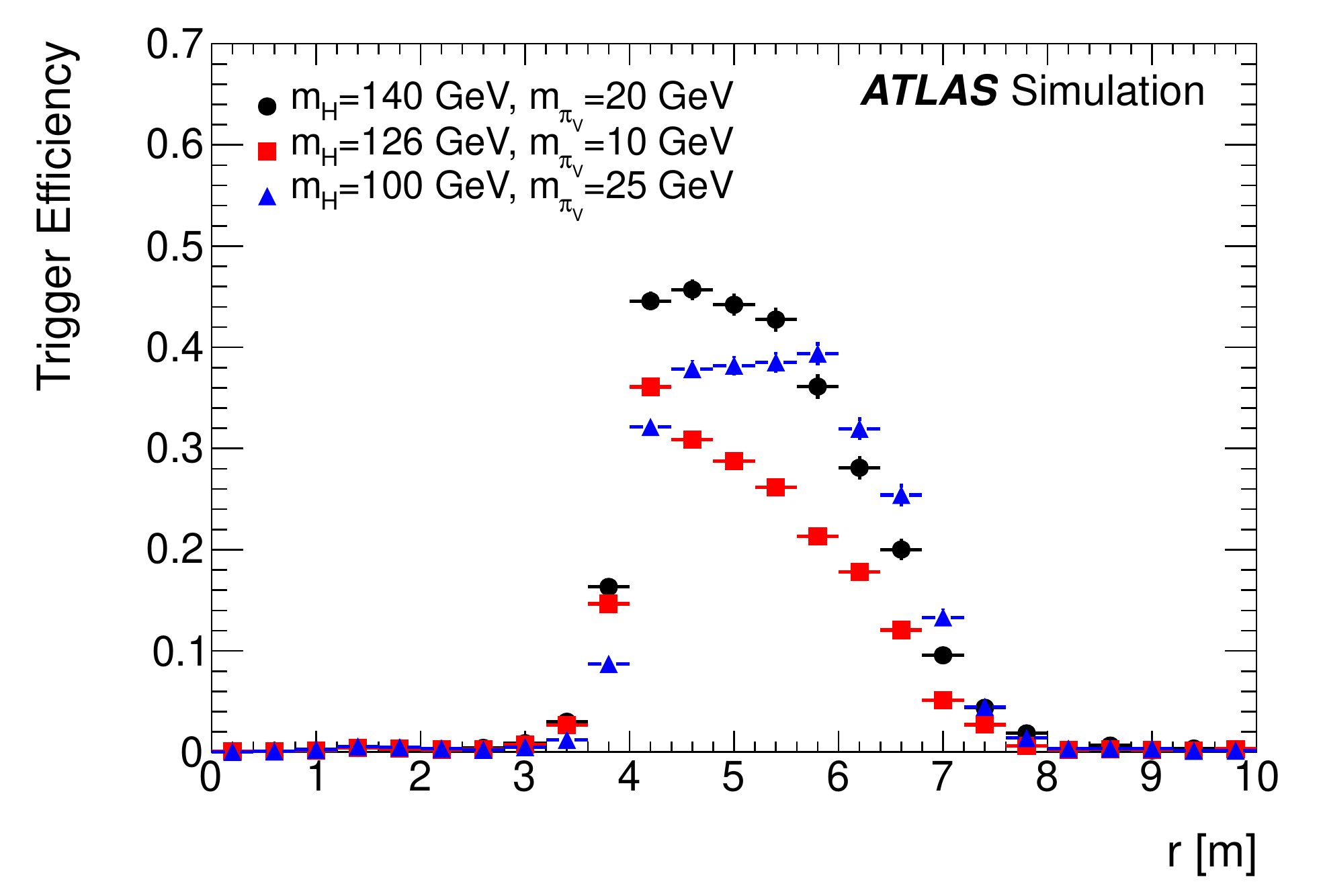}
	\caption{}
	\end{subfigure}
	\begin{subfigure}[b]{0.47\textwidth}
	\includegraphics[width=\textwidth]{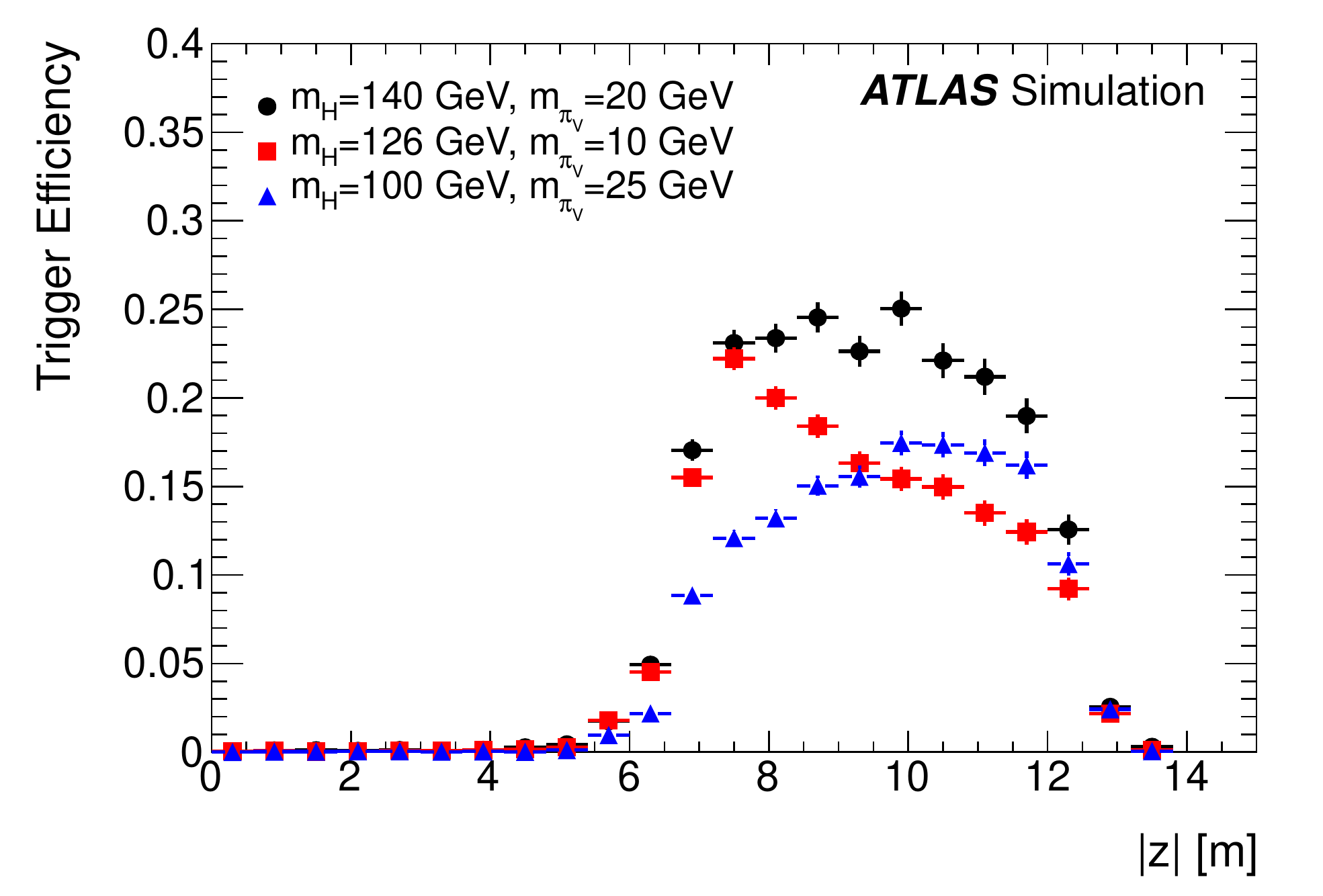}
	\caption{}
	\end{subfigure}
	\caption{Efficiency for the Muon \roi Cluster trigger as a function of (a) the radial decay position, $r$, for \piv decays in the barrel and (b) the $|z|$ position for \piv decays in the \Endcaps.}
	\label{fig:MuonClusterTriggerEff}
\end{figure}

The efficiency is defined as the fraction of \piv particles that pass one of the triggers for displaced decays of long-lived neutral particles as a function of the \piv decay position. Figures~\ref{fig:TracklessTriggerEff}--\ref{fig:MuonClusterTriggerEff} show the long-lived neutral particle trigger efficiencies for the simulated mass points, with mean $\langle\mu\rangle=22$, as a function of the \piv decay position for decays in the barrel and \Endcap regions of the ATLAS detector. The uncertainties shown are statistical only. The Calorimeter Ratio trigger and Muon RoI cluster triggers have good efficiency throughout the detector volume. All of the efficiencies are lower in the forward region than in the central region. In general, this is a consequence of events not satisfying the isolation criteria because of the higher mean occupancy in the forward region due to \pileup interactions. 

The Trackless Jet trigger is less than 5\% efficient for decays occurring in the region of the barrel between the middle of the ID and the \ECal. The trigger efficiency increases as the decay occurs closer to the end of the \HCal. The dependence of the efficiency on the decay position is a consequence of the low $\pT$ distribution of muons produced by the \piv decays. A muon originating in the ID will lose an average of 4\,GeV of energy as it traverses the ATLAS calorimeters \cite{CSCBOOK}; therefore, a larger fraction of muons produced in \piv decays reach the muon system as the \piv decays further into the calorimeter. The efficiency in the \Endcap regions is $\sim1$--2\%. The Calorimeter Ratio trigger is $\sim25$\% efficient for \piv decays in the \HCal barrel and 10\% in the \HCal \Endcaps. The Muon \roi Cluster trigger in the MS barrel is $\sim40$\% efficient from the end of the \HCal to $r\sim6$\,m. The efficiency drops beyond a radial distance of 6\,m because the jets originating from the \piv decay do not separate sufficiently before reaching the first muon trigger plane located at $r\sim7$\,m. The trigger efficiency in the MS \Endcaps is 15--25\% for $|z|$ between 7 and 12\,m from the IP.

The lower efficiency of the Calorimeter Ratio trigger for the sample with $m_{\rm{H}}=100$\,GeV and $m_{\pi_{\rm{v}}}=25$\,GeV is due to the tight L1 energy threshold and the fact that jets produced by the \piv decays in this MC sample have lower energies compared to the other samples. The lower efficiency of the Muon \roi Cluster trigger and rapid decrease with increasing decay distance for the sample with $m_{\rm{H}}=126$\,GeV and $m_{\pi_{\rm{v}}}=10$\,GeV is due to the higher \piv boost that results in collinear jets that are more likely to be contained in a single MS sector. Because the L1 muon firmware is limited to no more than two \rois per sector, many of these decays do not satisfy the trigger condition that requires three (four) or more \rois in the MS barrel (\Endcaps). The effect is minimized if the jet axis falls between two adjacent MS sectors.

To evaluate the effect of \pileup interactions on the trigger efficiency, the MC samples are arbitrarily divided into two subsamples, one with $\langle\mu\rangle>22$ and one with $\langle\mu\rangle\leq22$, with mean $\langle\mu\rangle$ values of 28.6 and 15.8, respectively. Tables~\ref{tab:PileUpEffect_Barrel} and \ref{tab:PileUpEffect_Endcap} give, for the three triggers, the ratio of Efficiency(28.6)/Efficiency(15.8) in the barrel and end-cap detectors.\footnote{The efficiencies used in the ratio are averaged over the $r$ and $|z|$ ranges given in tables~\ref{tab:PileUpEffect_Barrel} and \ref{tab:PileUpEffect_Endcap}, respectively.} Both the Trackless Jet and the Calorimeter Ratio triggers show significant reduction of efficiency with increasing \pileup in both the barrel and \Endcap regions. These two triggers are particularly sensitive to \pileup interactions because of the tight isolation criterion that requires no reconstructed tracks at L2 with $\pT>0.8$ and $\pT>1$\,GeV in a cone around the jet axis for the Trackless Jet and Calorimeter Ratio triggers, respectively. The Muon \roi Cluster trigger includes  a softer track isolation cut, no reconstructed tracks at L2 with $\pT>5$\,GeV around the muon \roi cluster centre, and is more robust against \pileup. No effect is observed in the barrel region while in the \Endcaps there is an approximate 7\% reduction. The isolation cuts represent a compromise between maximizing the signal acceptance and reducing the final output rate to a sustainable level.

\begin{table}[t!]
\begin{center}
\begin{tabular}{lcccl}
\hline
\small{Sample} & \small{Trackless Jet} & \small{Calorimeter Ratio} & \small{Muon \roi Cluster} & \small{Range [m]} \\
\hline
$m_{\rm{H}}/m_{\pi_{\rm{v}}}=100/25$\,GeV & $0.70\pm0.04$ & $0.48\pm0.04$ & $1.03\pm0.03$ & $0.4<r<4.0$ \\
$m_{\rm{H}}/m_{\pi_{\rm{v}}}=126/10$\,GeV & $0.76\pm0.04$ & $0.47\pm0.02$ & $0.95\pm0.03$ & $2.0<r<4.0$ \\
$m_{\rm{H}}/m_{\pi_{\rm{v}}}=140/20$\,GeV & $0.64\pm0.02$ & $0.50\pm0.02$ & $0.98\pm0.03$ & $4.0<r<7.2$ \\
\hline
\end{tabular}
\caption{The ratio of the trigger efficiencies in the barrel for mean $\langle\mu\rangle$ = 28.6 (high \pileup sample) to mean $\langle\mu\rangle$ = 15.8 (low \pileup sample) for the three triggers and the three benchmark samples. The uncertainties shown are statistical only.}
\label{tab:PileUpEffect_Barrel}
\vspace{0.5cm}
\begin{tabular}{lcccl}
\hline
\small{Sample} & \small{Trackless Jet} & \small{Calorimeter Ratio} & \small{Muon \roi Cluster} & \small{Range [m]} \\
\hline
$m_{\rm{H}}/m_{\pi_{\rm{v}}}=100/25$\,GeV & $0.32\pm0.07$ & $0.34\pm0.06$ & $0.93\pm0.02$ & $1.8<|z|<6.0$ \\
$m_{\rm{H}}/m_{\pi_{\rm{v}}}=126/10$\,GeV & $0.51\pm0.08$ & $0.35\pm0.03$ & $0.93\pm0.02$ & $4.2<|z|<6.0$ \\
$m_{\rm{H}}/m_{\pi_{\rm{v}}}=140/20$\,GeV & $0.39\pm0.04$ & $0.40\pm0.04$ & $0.92\pm0.02$ & $6.0<|z|<13.2$ \\
\hline
\end{tabular}
\caption{The ratio of the trigger efficiencies in the \Endcaps for mean $\langle\mu\rangle$ = 28.6 (high \pileup sample) to mean $\langle\mu\rangle$ = 15.8 (low \pileup sample) for the three triggers and the three benchmark samples. The uncertainties shown are statistical only.}
\label{tab:PileUpEffect_Endcap}
\end{center}
\end{table}

\begin{figure}[th!]
	\centering
	\includegraphics[width=0.7\textwidth]{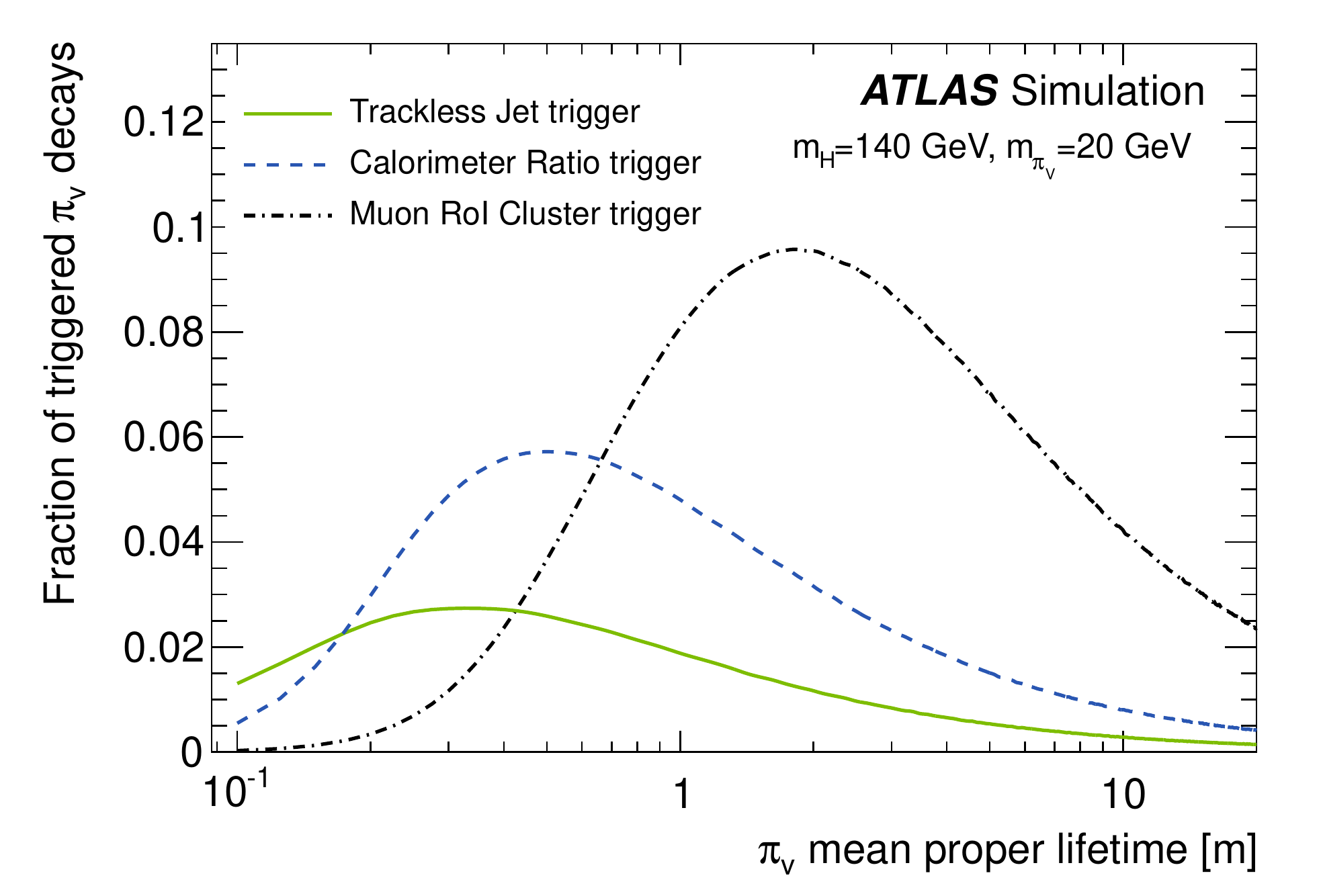}
	\caption{The expected fraction of events that pass each of the long-lived neutral particle triggers as a function of the \piv mean proper lifetime.}
	\label{fig:TriggerExpectedFraction}
\end{figure}

From the trigger efficiency plots, the signal fraction accepted by the long-lived neutral particle triggers is predicted as a function of the \piv mean proper lifetime. The fraction of \piv decays that occur at a certain radius is weighted by the corresponding trigger efficiency for the same radius. Figure~\ref{fig:TriggerExpectedFraction} shows the expected fraction of events that pass each trigger as a function of the \piv mean proper lifetime for $m_{\rm{H}}=140$\,GeV and $m_{\pi_{\rm{v}}}=20$\,GeV.

All the trigger studies and efficiencies presented in this paper are based on MC simulations and the associated uncertainties are statistical only. An analysis using one or more of these triggers will need to make a careful study of the systematic uncertainties. While the details of the study are analysis dependent, standard sources of systematic uncertainties include the jet energy scale, the jet energy resolution, the modelling of hard and soft proton-proton interactions and the instantaneous luminosity. Additional systematic uncertainty studies that focus on the physics objects used by these triggers need to be taken into account and some examples follow. 

The cluster of muon RoIs resulting from a \piv decay in the MS is the basic physics object used by the Muon \roi Cluster trigger. To compare the MS detector response to a \piv decay in data and simulation, a sample of jets that punch through the calorimeter can be used. The punch-through jets are similar to signal events in that both low-$\pT$ photons and charged hadrons are present in a narrow region of the MS, thus the detector response to such an environment can be verified between data and simulation. The systematic uncertainty related to the logarithmic ratio selection criteria can be studied by selecting a sample of jets with a small fraction of energy in the \ECal and comparing the distribution in data and simulation when applying the various cuts used in the Calorimeter Ratio trigger. For the Trackless Jet trigger, a data sample collected with a minimum bias trigger can be used to identify secondary vertex interactions that may mimic the decay topology of a \piv in the ID or \ECal.  In general, systematic uncertainties are analysis dependent and detailed studies are beyond the scope of this paper.

\section{Triggering on collision data}
\begin{figure}[!t]
	\centering
	\includegraphics[width=0.7\textwidth]{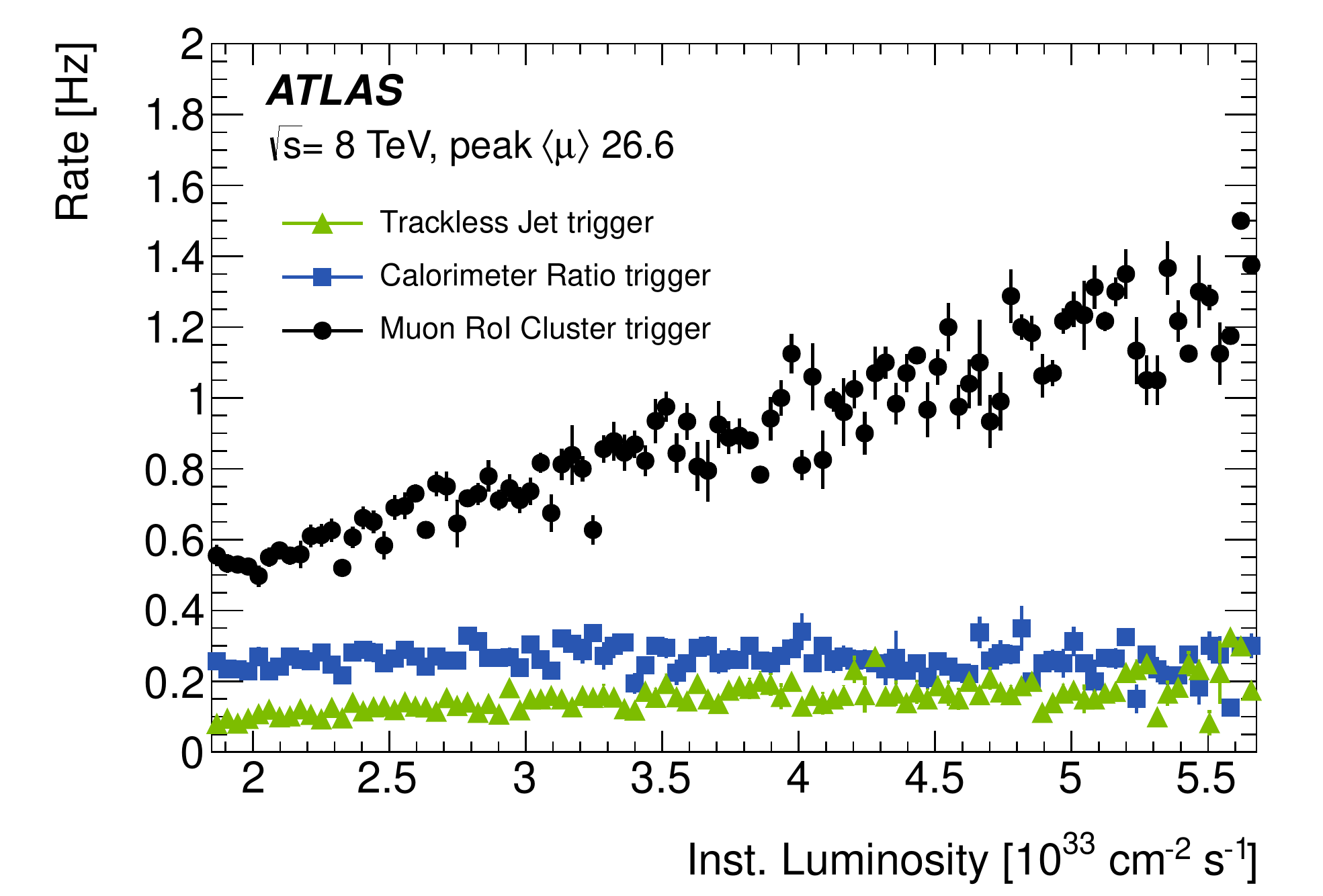}
	\caption{Trigger output rates as a function of the instantaneous luminosity during a typical 2012 data-taking period.}
	\label{fig:TriggerRate}
\end{figure}

These triggers need to meet tight bandwidth requirements for inclusion in the ATLAS trigger menu. The exclusive bandwidth allocated for the long-lived neutral particle triggers is only a few Hz. 
Figure~\ref{fig:TriggerRate} shows the rates as a function of instantaneous luminosity for the three triggers during a typical 2012 data-taking period and demonstrates that the rates are well within the allocated trigger bandwidth. The Muon \roi Cluster trigger shows a linear dependence on the instantaneous luminosity. A linear fit, when extrapolated to null luminosity and accounting for the different number of filled and empty bunches, predicts a rate of ($0.04\pm0.01$)\,Hz. An analogous trigger only enabled on empty bunch crossings gives a rate of ($0.04\pm0.03$)\,Hz in the same data-taking period as that displayed in figure~\ref{fig:TriggerRate}. The rates are found to be in excellent agreement, suggesting that the Muon \roi Cluster trigger rate increases linearly as a function of the instantaneous luminosity and there are no other dependencies affecting its rate. 

The Trackless Jet and Calorimeter Ratio trigger rates show little or no dependence on the instantaneous luminosity. As discussed in the previous section, this is the result of the tight track isolation requirement included among the \online selection cuts. The Muon \roi Cluster trigger uses a softer track isolation requirement and its rate is determined only by the instantaneous luminosity.

\begin{figure}[!t]
	\centering
	\includegraphics[width=0.7\textwidth]{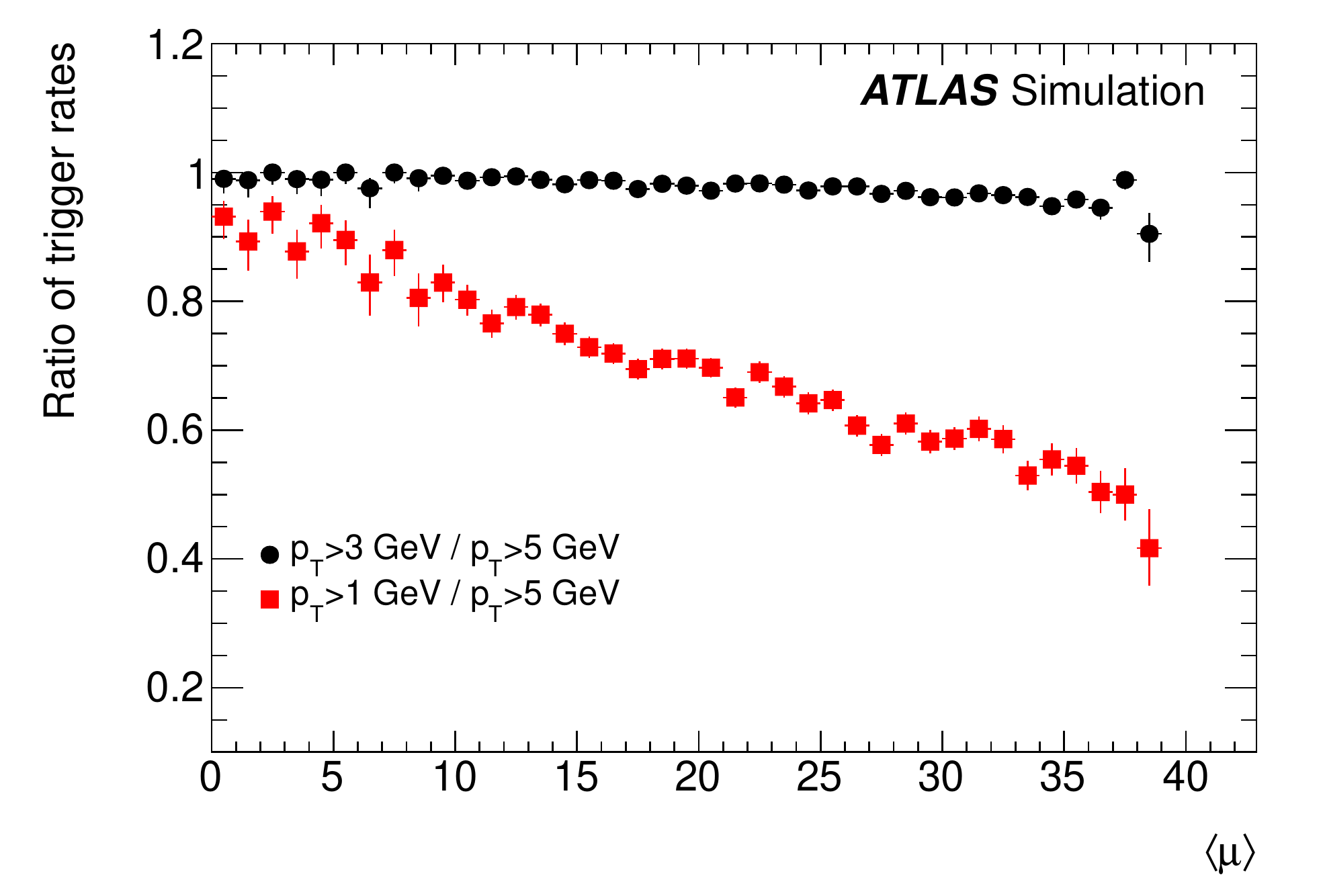}
	\caption{Ratio of trigger rates as a function of $\langle\mu\rangle$ when using lower $\pT$ isolation cuts for the Muon RoI Cluster trigger compared to the $\pT>5$\ GeV cut used online.}
	\label{fig:TriggerRatePileUp}
\end{figure}

More detailed studies on data are, in general, not possible since the instantaneous luminosity and \pileup are correlated quantities and the LHC running conditions do not provide different samples with the same instantaneous luminosity and different \pileup configurations.
However, the correlation between the track isolation criterion and the amount of \pileup in the event can be further studied using the Muon \roi Cluster trigger. Figure \ref{fig:TriggerRatePileUp} shows the ratio of the Muon \roi Cluster trigger rates as a function of $\langle\mu\rangle$ when lowering the threshold of the $\pT$ tracking isolation cut below the nominal $\pT>5$\,GeV used in the trigger. There is no difference in the rate when requiring isolation using tracks with $\pT>5$\,GeV or $\pT>3$\,GeV. However, a net linear rate reduction with increasing \pileup is observed when using tracks with $\pT>1$\,GeV. As expected, a higher number of interactions per event increases the likelihood that the event is rejected for triggers using low-momentum track isolation.

\section{Conclusions}
Three signature-driven triggers designed to select decays of long-lived neutral particles throughout the ATLAS detector volume are described: the Trackless Jet trigger for decays beyond the pixel layers to the \ECal, the Calorimeter Ratio trigger for decays in the \HCal and the Muon RoI Cluster trigger for decays from the end of the \HCal to the MS middle station. These triggers are demonstrated to be efficient for pairs of \piv particles produced in Higgs boson decays, using different lifetimes and a range of Higgs boson and \piv masses. They also met the tight bandwidth requirements of the ATLAS triggers during the 2012 LHC physics data-taking period. The estimated fraction of triggered \piv particles is between 2\% and 10\% for a \piv mean proper lifetime in the range 0.1\,m to 20\,m. The ATLAS physics potential for probing various SM extensions can be greatly extended using the triggers described in this paper.

% Acknowledgements for papers with collision data
% Version 1-Apr-2013

\section{Acknowledgements}

% Standard acknowledgements start here
%----------------------------------------------
We thank CERN for the very successful operation of the LHC, as well as the
support staff from our institutions without whom ATLAS could not be
operated efficiently.

We acknowledge the support of ANPCyT, Argentina; YerPhI, Armenia; ARC,
Australia; BMWF and FWF, Austria; ANAS, Azerbaijan; SSTC, Belarus; CNPq and FAPESP,
Brazil; NSERC, NRC and CFI, Canada; CERN; CONICYT, Chile; CAS, MOST and NSFC,
China; COLCIENCIAS, Colombia; MSMT CR, MPO CR and VSC CR, Czech Republic;
DNRF, DNSRC and Lundbeck Foundation, Denmark; EPLANET, ERC and NSRF, European Union;
IN2P3-CNRS, CEA-DSM/IRFU, France; GNSF, Georgia; BMBF, DFG, HGF, MPG and AvH
Foundation, Germany; GSRT and NSRF, Greece; ISF, MINERVA, GIF, DIP and Benoziyo Center,
Israel; INFN, Italy; MEXT and JSPS, Japan; CNRST, Morocco; FOM and NWO,
Netherlands; BRF and RCN, Norway; MNiSW, Poland; GRICES and FCT, Portugal; MERYS
(MECTS), Romania; MES of Russia and ROSATOM, Russian Federation; JINR; MSTD,
Serbia; MSSR, Slovakia; ARRS and MIZ\v{S}, Slovenia; DST/NRF, South Africa;
MICINN, Spain; SRC and Wallenberg Foundation, Sweden; SER, SNSF and Cantons of
Bern and Geneva, Switzerland; NSC, Taiwan; TAEK, Turkey; STFC, the Royal
Society and Leverhulme Trust, United Kingdom; DOE and NSF, United States of
America.

The crucial computing support from all WLCG partners is acknowledged
gratefully, in particular from CERN and the ATLAS Tier-1 facilities at
TRIUMF (Canada), NDGF (Denmark, Norway, Sweden), CC-IN2P3 (France),
KIT/GridKA (Germany), INFN-CNAF (Italy), NL-T1 (Netherlands), PIC (Spain),
ASGC (Taiwan), RAL (UK) and BNL (USA) and in the Tier-2 facilities
worldwide.

\onecolumn
\clearpage
% ATLAS Collaboration author list for 12-FEB-2013
% Data extracted on 19-Mar-2013 for paperid 332
%\documentclass[11pt]{article}
%\usepackage{a4wide}\begin{document}
\begin{flushleft}
{\Large The ATLAS Collaboration}

\bigskip

G.~Aad$^{\rm 48}$,
T.~Abajyan$^{\rm 21}$,
B.~Abbott$^{\rm 112}$,
J.~Abdallah$^{\rm 12}$,
S.~Abdel~Khalek$^{\rm 116}$,
A.A.~Abdelalim$^{\rm 49}$,
O.~Abdinov$^{\rm 11}$,
R.~Aben$^{\rm 106}$,
B.~Abi$^{\rm 113}$,
M.~Abolins$^{\rm 89}$
% ...
\bigskip

% ...
$^{11}$ Institute of Physics, Azerbaijan Academy of Sciences, Baku, Azerbaijan\\
$^{12}$ Institut de F\'isica d'Altes Energies and Departament de F\'isica de la Universitat Aut\`onoma de Barcelona and ICREA, Barcelona, Spain\\
% ...
$^{21}$ Physikalisches Institut, University of Bonn, Bonn, Germany\\
% ...
$^{48}$ Fakult\"{a}t f\"{u}r Mathematik und Physik, Albert-Ludwigs-Universit\"{a}t, Freiburg, Germany\\
$^{49}$ Section de Physique, Universit\'e de Gen\`eve, Geneva, Switzerland\\
% ...
$^{89}$ Department of Physics and Astronomy, Michigan State University, East Lansing MI, United States of America\\
% ...
$^{106}$ Nikhef National Institute for Subatomic Physics and University of Amsterdam, Amsterdam, Netherlands\\
% ...
$^{112}$ Homer L. Dodge Department of Physics and Astronomy, University of Oklahoma, Norman OK, United States of America\\
$^{113}$ Department of Physics, Oklahoma State University, Stillwater OK, United States of America\\
% ...
$^{116}$ LAL, Universit\'e Paris-Sud and CNRS/IN2P3, Orsay, France\\
% ...
\end{flushleft}

%\end{document}


\begin{thebibliography}{}
\bibitem{SUSY} S.\,Dimopoulos, M.\,Dine, S.\,Raby and S.\,Thomas, \emph{Experimental signatures of low energy gauge-mediated supersymmetry}, \emph{Phys. Rev. Lett.} {\bf 76} (1996) 3494.
\bibitem{MSSM} R. Barbier et al., \emph{R-parity violating supersymmetry}, \emph{Phys. Rept.} {\bf 420} (2005) 1.
\bibitem{DarkMatter} D.\,Smith and N.\,Weiner, \emph{Inelastic dark matter}, \emph{Phys. Rev.} {\bf D64} (2001) 043502.
\bibitem{HV1} M. J. Strassler and K. M. Zurek, \emph{Echoes of a hidden valley at hadron colliders}, \emph{Phys. Lett.}  {\bf B651} (2007) 374.
\bibitem{HV2} M. J. Strassler and K. M. Zurek, \emph{Discovering the Higgs through highly-displaced vertices}, \emph{Phys. Lett.} {\bf B661} (2008) 263.
\bibitem{HV3} M. J. Strassler, \emph{Possible effects of a hidden valley on a supersymmetric phenomenology}, hep-ph/0607160.
\bibitem{HVAtlas} ATLAS Collaboration, \emph{Search for a Light Higgs Boson Decaying to Long-Lived Weakly Interacting Particles in Proton-Proton Collisions at $\sqrt{s}=7$\,TeV with the ATLAS Detector}, \emph{Phys. Rev. Lett.} {\bf 108} (2012) 251801.
\bibitem{ATLASTDR}{ATLAS Collaboration}, \emph{The ATLAS Experiment at the CERN Large Hadron Collider}, \emph{JINST} {\bf 3} (2008)  S08003.
\bibitem{TRIG}{ATLAS Collaboration}, \emph{Performance of the ATLAS Trigger System in 2010}, \emph{Eur. Phys. J.} {\bf C72} (2012) 1849.
\bibitem{PYTHIA8} S.\,Mrenna, T.\,Sjostrand and P.\,Z.\,Skands, \emph{A brief introduction to PYTHIA 8.1}, \emph{Comput. Phys. Commun.} {\bf 178} (2008) 852.
\bibitem{PDF} A. Martin, W. Stirling, R. Thorne and G. Watt, \emph{Parton distributions for the LHC}, \emph{Eur. Phys. J.} {\bf C63} (2009) 189.
\bibitem{OFA} E. Fullana et al., \emph{Digital Signal Reconstruction in the ATLAS Hadronic Tile Calorimeter}, \emph{IEEE Trans. Nucl. Sci.} {\bf 53} (2006) 4.
\bibitem{HCalSignalReconstruction} A. Valero, \emph{Implementation and Performance of the Signal Reconstruction in the ATLAS Hadronic Tile Calorimeter}, ATL-TILECAL-PROC-2011-008.
\bibitem{AntiKt} M. Cacciari, G. P. Salam and G. Soyez, \emph{The anti-$k_t$ jet clustering algorithm}, \emph{JHEP} {\bf 04} (2008) 063.
\bibitem{H8TB} C. Adorisio et al., \emph{System test of the ATLAS muon spectrometer in the H8 beam at the CERN SPS}, \emph{Nucl. Instrum. Meth.} {\bf A593} (2008) 232.
\bibitem{CSCBOOK}ATLAS Collaboration, \emph{Expected Performance of the ATLAS Experiment - Detector, Trigger and Physics}, hep-ph/0901.0512.
\end{thebibliography}
\end{document}